\documentclass[aps,prl,reprint,superscriptaddress,letterpaper,twocolumn,longbibliography,floatfix]{revtex4-2}

\usepackage{polyglossia}
\setmainlanguage{english}

\usepackage{graphicx}
\usepackage[svgnames,dvipsnames,x11names]{xcolor}
\usepackage[colorlinks=true,linkcolor=SpringGreen4,citecolor=blue,urlcolor=Magenta3]{hyperref}

\usepackage{amsmath}
\usepackage{amssymb}
\usepackage{float}
\usepackage{amsthm}
\usepackage{amsfonts}
\usepackage{mathtools}
\usepackage{mathrsfs}
\usepackage{bm}
\usepackage{bbm}
\usepackage{empheq}
\usepackage{cases}
\usepackage{euscript}

\usepackage[draft,todonotes={textsize=scriptsize}]{changes}
\setuptodonotes{inline}
\definechangesauthor[name={Nada Khogeer}, color=Blue4]{NK}
\definechangesauthor[name={Jerison Parra}, color=Purple]{JP}
\definechangesauthor[name={Hugo Ribeiro}, color=SpringGreen4]{HR}

\newcommand{\bra}[1]{\langle #1|}
\newcommand{\ket}[1]{|#1\rangle}

\newcommand{\ketbra}[2]{\ket{#1}\!\bra{#2}}
\newcommand{\norm}[1]{\left\lVert#1\right\rVert}
\newcommand{\mm}[1]{\mathrm{#1}}
\newcommand{\abs}[1]{\left|#1\right|}

\newcommand{\di}[1]{\mathop{}\!\mathrm{d} #1}
\newcommand{\com}[2]{\left[#1,#2\right]}
\newcommand{\avg}[1]{\langle #1 \rangle}

\def \Id{\mathbbm{1}}
\def \rd{\partial}
\DeclareMathOperator{\Tr}{Tr}
\DeclareMathOperator{\Ci}{Ci}
\DeclareMathOperator{\ad}{ad}
\DeclareMathOperator{\sinc}{sinc}
\DeclareMathOperator{\BigO}{\mathcal{O}}

\def \uc{\mathrm{c}}
\def \ud{\mathrm{d}}

\def \ug{\mathrm{g}}

\def \uq{\mathrm{q}}

\def \uq{\mathrm{q}}

\def \uG{\mathrm{G}}
\def \uI{\mathrm{I}}

\def \rv{\mbox{\boldmath$r$}}

\def \hrho{\hat{\rho}}
\def \hsigma{\hat{\sigma}}

\def \hA{\hat{A}}
\def \hB{\hat{B}}

\def \hH{\hat{H}}

\def \hO{\hat{O}}

\def \hT{\hat{T}}
\def \hU{\hat{U}}
\def \hV{\hat{V}}

\def \sL{\EuScript{L}}
\def \slr{\mathscr{L}}

\begin{document}

\title{Novikov-Magnus Master Equation for Driven Quantum Systems under Classical Correlated Noise}

\author{Nada Khogeer}
\affiliation{Department of Physics and Applied Physics, University of Massachusetts Lowell, Lowell, MA 01854, USA}

\author{Jerison Parra}
\affiliation{Department of Physics and Applied Physics, University of Massachusetts Lowell, Lowell, MA 01854, USA}

\author{Caroline Desgranges}
\affiliation{Department of Physics and Applied Physics, University of Massachusetts Lowell, Lowell, MA 01854, USA}

\author{Hugo Ribeiro}
\affiliation{Department of Physics and Applied Physics, University of Massachusetts Lowell, Lowell, MA 01854, USA}

\begin{abstract}
We introduce a modular framework for deriving time-local quantum master equations for driven systems subject to temporally
correlated classical noise. The construction combines the Donsker--Furutsu--Novikov relation with the Magnus expansion:~Functional
response closes the noise average, while nested commutators retain the noncommuting chronological action of noise-induced
perturbations along the controlled quantum trajectory. For zero-mean Gaussian noise, we derive the resulting Novikov--Magnus
quantum master equation through fourth order in the noise strength. The same response hierarchy suggests a natural extension
beyond Gaussian noise, with connected higher-order noise correlations weighting higher functional derivatives of the stochastic
map. For a driven qubit under Ornstein--Uhlenbeck and band-limited $1/f$ noise, the second- and fourth-order maps remain CPTP
throughout the perturbative regime and accurately reproduce stochastic simulations. Their diamond-norm errors are comparable to
those of the cumulant-based pseudo-Lindblad master equation, with a modest fourth-order improvement for Ornstein--Uhlenbeck noise.
\end{abstract}

\maketitle

\begin{figure}[t]
    \includegraphics[width=0.99\columnwidth]{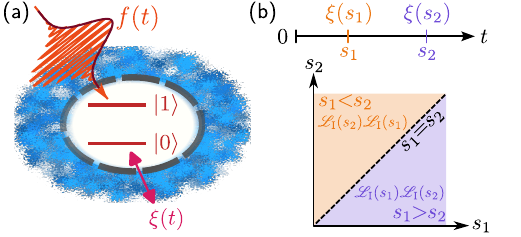}
    \caption{
        (a) A driven qubit subject to temporally correlated classical noise. The control field $f(t)$ generates the target gate
        and rotates the interaction-picture noise-coupling operator. (b) The Donsker--Furutsu--Novikov relation introduces
        statistically weighted response times, while the corresponding quantum actions $\slr_\uI(s_i)$ generally do not commute.
        Resolving the integration domain into chronological sectors associates each ordering of the insertion times with the
        corresponding ordered quantum response.
    }
    \label{fig:SystemBath}
\end{figure}

\paragraph{Introduction.---}
Temporally correlated noise limits the control of solid-state quantum systems. As illustrated in Fig.~\ref{fig:SystemBath}(a), a
driven system evolves under both an applied control and environmental fluctuations. When these fluctuations persist over a gate,
the control rotates the operator through which the noise couples to the system. Noise acting at different times can then generate
noncommuting quantum operations. The resulting dynamics depend on the complete noise history and cannot generally be described by
independent errors or by dephasing rates calculated without the drive ~\cite{Ithier2005,Paladino2014,Hangleiter2021,Nakamura2024}.

Standard Lindblad master equations provide a convenient time-local description when environmental correlations decay rapidly
compared with the system dynamics and the Markov approximation is justified~\cite{Gorini1976,Lindblad1976,heinz07}. For
persistent correlations, the Nakajima--Zwanzig equation is nonlocal in time, and time-convolutionless expansions become
increasingly complicated at higher orders. Hierarchical equations of motion retain environmental memory through coupled auxiliary
operators and can access strongly non-Markovian regimes, but require propagation and truncation of an extended
hierarchy~\cite{Nakajima1958,Zwanzig1960,Shibata1977,Chaturvedi1979,Ishizaki2009}. Direct stochastic propagation instead requires
averaging over many noise realizations, while filter-function methods do not generally produce a compact time-local master
equation~\cite{Hangleiter2021,Chalermpusitarak2021}.

For Gaussian classical noise, Groszkowski~\textit{et al.} derived a pseudo-Lindblad master equation (PLME) from a generalized
cumulant expansion~\cite{Groszkowski_2023}. The PLME captures the noncommuting drive--noise interplay through fourth order and
accurately describes driven-qubit dynamics under several noise spectra. We therefore develop and benchmark the present closure for
zero-mean Gaussian noise, for which the PLME provides the closest time-local comparison.

The remaining problem is to derive higher-order time-local master equations without directly expanding and reorganizing the
averaged propagator. We solve this closure problem with the Novikov--Magnus quantum master equation (NMQME). For zero-mean
Gaussian noise, the Donsker--Furutsu--Novikov relation replaces correlations between the noise and the stochastic evolution by
functional responses ~\cite{Donsker1964,Furutsu1963,Novikov1965}. The Magnus expansion then expresses each response as a
superoperator acting on the stochastic map and preserves its noncommuting chronological structure through nested
commutators~\cite{Magnus1954,Blanes2009}. Recursive averaging produces a systematic hierarchy of time-local terms, which we
evaluate through fourth order.

This organization separates two ingredients that are mixed in a direct propagator expansion:~The noise statistics determine the
scalar weights of the response insertions, while the controlled dynamics determine their ordered quantum action. Changing the
Gaussian correlation function or control waveform requires reevaluating the corresponding kernels and commutators, but not
rederiving the closure procedure. Generalized functional identities extend the same procedure to non-Gaussian classical
noise~\cite{AthanassoulisMamis2019}, and an operator-valued response identity may provide an analogous route for quantum
environments.

We test the NMQME at the level of complete finite-time channels for a driven qubit under Ornstein--Uhlenbeck and band-limited
$1/f$ noise. Choi-matrix diagnostics test whether the approximate maps are completely positive and trace preserving, while the
diamond norm measures their worst-case error relative to stochastic simulations. Throughout the perturbative regime, the NMQME
remains physical and matches the accuracy of the PLME. At fourth order, it gives a modest improvement for OU noise, while the two
methods are quantitatively equivalent for band-limited $1/f$ noise. Detailed derivations, additional dynamical benchmarks, and
validation of the stochastic noise generation are provided in the Supplemental Material~\cite{SupplementalMaterial}.

\paragraph{Noise-averaged stochastic dynamics.---}
We consider a quantum system governed by the stochastic Hamiltonian
\begin{equation}
	\hH (t) = \hH_0 (t) + \varepsilon \xi (t) \hV (t),
	\label{eq:fullH}
\end{equation}
where $\hH_0(t)$ describes the noise-free dynamics, $\hV(t)$ is the operator through which the noise couples to the system, and
$\xi(t)$ is a real-valued classical stochastic process. The parameter $\varepsilon$ is introduced to organize the perturbative
expansion and is set to one after the expansion is truncated. Our goal is to derive an equation of motion for the dynamics
averaged over realizations of $\xi(t)$.

In the absence of noise, $\hH_0(t)$ generates the unitary evolution operator
\begin{equation}
	\hU_0 (t) = \hT \exp\left[ -i \int_0^{t} \di{t_1}\, \hH_0(t_1) \right],
	\label{eq:U0}
\end{equation}
where $\hT$ denotes time ordering and we set $\hbar=1$. The Hamiltonian $\hH_0 (t)$ is chosen such that at $t=t_\ug$ we have $\hU_0 (t_\ug) =
\hU_\uG$, with $\hU_\uG$ a desired target unitary gate.

We assume that $\xi(t)$ is a zero-mean Gaussian process characterized by the two-time correlation function
\begin{equation} 
    C(t,s) = \avg{\xi(t)\xi(s)}_\xi, 
    \label{eq:NoiseCorrelator} 
\end{equation} 
where $\avg{\cdot}_\xi$ denotes an average over noise realizations. We do not assume that the noise is Markovian, so $C(t,s)$ may
retain correlations over times comparable to the system evolution. Because the process is Gaussian, all of its higher-order
moments are determined by $C(t,s)$.


To isolate the effect of the noise on the dynamics, we move to the interaction picture generated by $\hH_0(t)$. In this frame, the
Hamiltonian becomes
\begin{equation} 
    \hH_\uI(t) = \hU_0^\dagger(t)\hH(t)\hU_0(t) -i\hU_0^\dagger(t)\rd_t\hU_0(t) = \varepsilon \xi (t) \hV_\uI (t),
\end{equation}
where $\hV_\uI(t) = \hU_0^\dagger(t) \hV(t) \hU_0(t)$ and we used $i\rd_t\hU_0(t)=\hH_0(t)\hU_0(t)$. The interaction-picture
dynamics are therefore generated entirely by the stochastic contribution, while the deterministic control remains encoded in the
time dependence of $\hV_\uI(t)$.

For a fixed realization of the noise, the interaction-picture density operator obeys 
\begin{equation} 
    \rd_t \hrho_\uI(t) = -i\com{\hH_\uI(t)}{\hrho_\uI(t)} = \varepsilon\,\xi(t)\, \slr_\uI(t)\hrho_\uI(t), 
    \label{eq:LiouvilleInteraction} 
\end{equation} 
where $\slr_\uI(t) \hO = -i\com{\hV_\uI(t)}{\hO}$. To describe this evolution independently of the initial state, we introduce the
stochastic dynamical map $\sL_\uI (t)$ through $\hrho_\uI (t) =\sL_\uI (t) \hrho(0)$. Equation~\eqref{eq:LiouvilleInteraction} then implies 
\begin{equation} 
    \rd_t\sL_\uI(t) = \varepsilon\,\xi(t)\, \slr_\uI(t)\sL_\uI(t), 
\label{eq:LI} 
\end{equation}
with initial condition $\sL_\uI(0) = \Id$. For each noise realization, Eq.~\eqref{eq:LI} determines a different dynamical map. The
physical evolution in the presence of classical noise is obtained by averaging these maps over the noise ensemble.

Averaging Eq.~\eqref{eq:LI} over noise realizations gives 
\begin{equation} 
    \rd_t\avg{\sL_\uI(t)}_\xi = \varepsilon\,\slr_\uI(t) \avg{\xi(t)\sL_\uI [\xi (t)]}_\xi, 
    \label{eq:AvgLI} 
\end{equation} 
where the notation $\sL_\uI [\xi (t)]$ emphasizes that the stochastic dynamical map is a functional of the noise history. Our
objective is to express the right-hand side solely in terms of the averaged map $\avg{\sL_\uI [\xi (t)]}_\xi$, thereby obtaining a closed
master equation. The difficulty is that $\sL_\uI [\xi (t)]$ is generated by the same noise realization that appears explicitly in
Eq.~\eqref{eq:AvgLI}. Thus, the mixed average $\avg{\xi(t)\sL_\uI [\xi (t)]}_\xi$ cannot be factorized.

For a zero-mean Gaussian process, the Donsker--Furutsu--Novikov relation expresses the correlation between the noise and a
functional of its history in terms of the functional response weighted by the covariance~\cite{Donsker1964,Furutsu1963,Novikov1965},
\begin{equation}
    \avg{\xi(t)\sL_\uI [\xi (t)]}_\xi = \int_0^t \di{s} C(t,s) \left\langle \frac{\delta \sL_\uI [\xi(t)]} {\delta\xi(s)} \right\rangle_\xi.
    \label{eq:NovikovZeroMeanGaussian}
\end{equation}
This relation reduces the closure problem to evaluating $\delta\sL_\uI[\xi(t)]/\delta\xi(s)$ and relating it to the stochastic map.

The Gaussian assumption is not intrinsic to the functional-response closure. For a general zero-mean process, the
noise--functional correlation admits the cumulant expansion
\begin{equation}
    \begin{aligned}
        \avg{\xi(t)F[\xi]}_\xi &= \sum_{n=1}^{\infty}\frac{1}{n!} \prod_{j=1}^{n}\int_0^t\di{s_j} C_{n+1}(t,s_1,\ldots,s_n) \\
        &\phantom{={}} \times \left\langle \frac{\delta^nF[\xi]} {\delta\xi(s_1)\cdots\delta\xi(s_n)} \right\rangle_\xi.
    \end{aligned}
    \label{eq:NovikovCumulantExpansion}
\end{equation}
where $C_n$ is the connected $n$-point noise correlation. This expansion separates the noise statistics from the dynamical
response:~non-Gaussianity enters through higher connected correlations, which weight higher functional derivatives. We focus here
on zero-mean Gaussian noise because its statistics are completely determined by the two-point correlation function $C(t,s)$. This
allows us to isolate the remaining challenges of closing the averaged dynamics and retaining the noncommuting time ordering
generated by the control. The Gaussian case also enables a controlled fourth-order comparison with the PLME and direct
benchmarking against stochastic simulations.

\paragraph{Magnus closure of the averaged dynamics.---} 
Equation~\eqref{eq:NovikovCumulantExpansion} organizes the statistical weights of the functional-response hierarchy. We now use
the Magnus representation to evaluate these responses while retaining the noncommuting chronological structure of the controlled
dynamics. The formal solution of Eq.~\eqref{eq:LI} is a time-ordered exponential whose functional dependence on the noise history
is difficult to evaluate directly. We instead write the stochastic map in the Magnus form
\begin{equation}
    \sL_\uI [\xi(t)] = \exp\left\{ \Omega[\xi (t)] \right\},
	\label{eq:LIMagnus}
\end{equation}
where $\Omega [\xi(t)] = \sum_{k=1}^\infty \varepsilon^k \Omega_k[\xi(t)]$, and $\Omega_k[\xi(t)]$ is the $k$th term of the Magnus
expansion~\cite{Magnus1954,Blanes2009}. Each term $\Omega_k [\xi(t)]$ contains $k$ factors of the stochastic process, so
Eq.~\eqref{eq:LIMagnus} organizes both the noise dependence and the functional response perturbatively. At the same time, the
nested commutators appearing in the Magnus terms account for the time ordering of the evolution. 

The functional derivative of Eq.~\eqref{eq:LIMagnus} can then be written as
\begin{equation}
    \frac{\delta \sL_\uI[\xi(t)]}{\delta \xi(s)} = \sum_{k=0}^\infty \frac{1}{(k+1)!} \ad^k_{\Omega[\xi (t)]}\left[
        \frac{\delta \Omega[\xi(t)]}{ \delta \xi(s)}\right] \sL_\uI [\xi(t)],
	\label{eq:FunDerLIMagnusFinal}
\end{equation}
where $\ad^0_A (B) = B$ and $\ad_A^k(B) = [A,[A,\ldots,[A,B]\ldots]]$, with $k$ nested commutators.
Equation~\eqref{eq:FunDerLIMagnusFinal} is the central closure step. Magnus time ordering converts the functional derivative into
a response superoperator acting on the stochastic map, allowing the remaining noise dependence to be averaged recursively.

Substituting Eq.~\eqref{eq:FunDerLIMagnusFinal} into Eq.~\eqref{eq:NovikovZeroMeanGaussian} does not immediately remove all noise
dependence because the Magnus exponent and its response remain functionals of $\xi(t)$. We therefore expand the response
superoperator in $\varepsilon$ and repeatedly apply the Donsker--Furutsu--Novikov relation to the remaining mixed averages.
Explicit noise factors are thereby replaced by correlation functions and additional functional derivatives.

A central feature of this modular construction is that classical noise can generate a noncommuting quantum response. The
Donsker--Furutsu--Novikov relation represents noise correlations as causal functional-response insertions at earlier times along
the quantum trajectory, consistent with statistical response theory~\cite{kubo1957,martin1973}. Although the noise amplitudes are
classical, the corresponding quantum actions are generated by $\slr_\uI(s_i)$. Since the control rotates the noise-coupling
operator during the evolution, these actions do not necessarily commute, and their chronological order matters. The Magnus nested
commutators retain this ordered response. At higher orders, recursive averaging introduces independently integrated insertion
times whose relative order is not fixed. The resulting partially ordered domains must therefore be decomposed into all
chronological sectors compatible with the ordering constraints already imposed by the Magnus integrals~\cite{Blanes2009}, as
illustrated in Fig.~\ref{fig:SystemBath}(b) and detailed in the Supplemental Material~\cite{SupplementalMaterial}.

After this recursive averaging and chronological decomposition, each perturbative order becomes a time-local superoperator acting
on the averaged map. Retaining all contributions through fourth order in $\varepsilon$ yields
\begin{equation}
    \partial_t \avg{\sL_\uI (t)}_\xi = \left[ \Xi^{(2)} (t) + \Xi^{(4)} (t)\right]\avg{\sL_\uI (t)}_\xi +
    \BigO\left(\varepsilon^6\right).
    \label{eq:finalEOMKernel}
\end{equation}
Because the noise is Gaussian with zero mean, odd-order contributions vanish, leaving $\Xi^{(2)}(t)=\BigO(\varepsilon^2)$ and
$\Xi^{(4)}(t)=\BigO(\varepsilon^4)$. Equation~\eqref{eq:finalEOMKernel} is time local, while its kernels retain the temporal noise
correlations. Each $\Xi^{(n)}(t)$ combines an $n$th-order statistical weight with the corresponding Magnus-ordered response.

\begin{figure*}[t] 
    \centering 
    \includegraphics[width=0.99\textwidth]{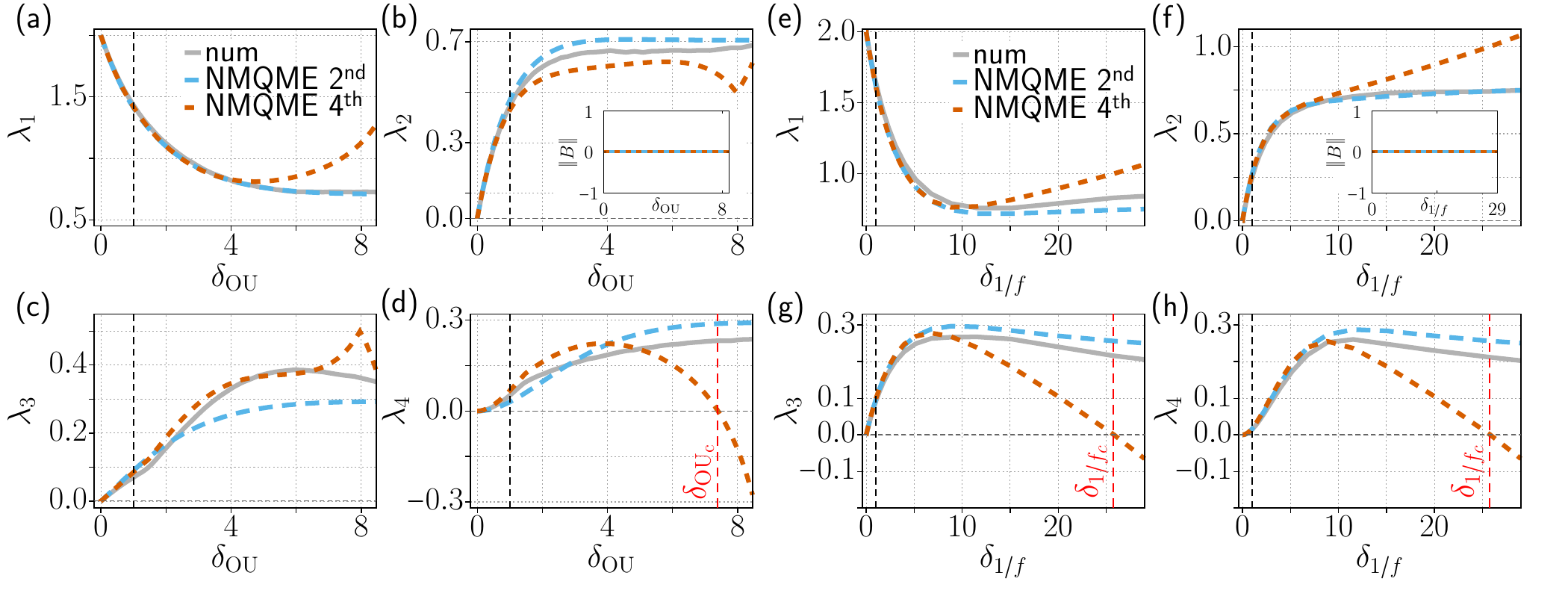}
    \caption{
        Choi eigenvalues for OU noise at $t_\ug=1.5\tau_\uc$ [(a)--(d)] and band-limited $1/f$ noise [(e)--(h)]. Stochastic
        simulations (gray) are compared with the second-order (blue dashed) and fourth-order (orange dashed) NMQME. Black dashed lines
        mark $\delta=1$, and red dashed lines mark the loss of fourth-order complete positivity. Insets in (b) and (f) show the
        trace-preservation residual.
    } 
    \label{fig:SemiDefPositiveChoiAll}
\end{figure*}

\paragraph{Physicality of the approximate dynamics.---}
The NMQME generator is derived as a perturbative series and retained only through a finite order in the noise strength. Although
the resulting equation is accurate to that order, the propagated finite-time map is not automatically guaranteed to be completely
positive and trace preserving. We therefore test these properties using the Choi matrix of the Schrödinger-picture map
\begin{equation}
    \avg{\sL(t)}_\xi = \sL_0(t)\avg{\sL_\uI(t)}_\xi,
    \label{eq:SchrodingerPictureMap}
\end{equation}
where $\sL_0(t)\hO=\hU_0(t)\hO\hU_0^\dagger(t)$. Since $\sL_0(t)$ is unitary, physicality may be assessed in either picture. For
the unnormalized Choi matrix $C_\Phi$, complete positivity and trace preservation are equivalent to
~\cite{Jamiolkowski1972,Choi1975,Wilde_2017}
\begin{equation}
    \begin{aligned}
        C_\Phi &\succeq 0, \\
        \Tr_\mm{out}[C_\Phi] &= \Id_\mm{in},
    \end{aligned}
    \label{eq:ChoiConditions}
\end{equation}
respectively. At $t=t_\ug$ we monitor
\begin{equation}
    \begin{aligned}
        \lambda_\mm{min}(t_\ug) &= \min_l\lambda_l[C_\Phi(t_\ug)],\\
        \Delta_\mm{TP}(t_\ug) &= \norm{\Tr_\mm{out}[C_\Phi(t_\ug)]-\Id_\mm{in}}_1.
    \end{aligned}
    \label{eq:MinChoiEigvalAndTPResidual}
\end{equation}
The map is CPTP when $\lambda_\mm{min}\geq0$ and $\Delta_\mm{TP}=0$ within numerical precision. Choi-matrix conventions and
physicality diagnostics are given in the Supplemental Material~\cite{SupplementalMaterial}.

\paragraph{Channel-distance benchmark.---}
Physical admissibility does not quantify accuracy. We compare the NMQME channel $\Phi_\mm{A}(t_\ug)$ with a reference channel
$\Phi_\mm{E}(t_\ug)$ obtained from stochastic simulations. Their worst-case channel distance is quantified by the diamond norm
~\cite{Aharonov1998,Wilde_2017,Watrous2018},
\begin{equation}
    \begin{aligned} 
        d_\diamond(t_\ug) &\coloneq \norm{\Phi_\mm{E}(t_\ug)-\Phi_\mm{A}(t_\ug)}_\diamond \\ 
        &= \max_{\hrho_{\mm{in},\mm{a}}} \norm{ \left[ \left(\Phi_\mm{E}-\Phi_\mm{A}\right) \otimes\mathcal{I}_\mm{a} \right]
        \left(\hrho_{\mm{in},\mm{a}}\right) }_1.
    \end{aligned}
    \label{eq:DiamondNormDifference}
\end{equation}
Here, $\norm{X}_1=\Tr(\sqrt{X^\dagger X})$ is the trace norm, which quantifies the distinguishability of the two output states, and
the maximization is over density operators $\hrho_{\mm{in},\mm{a}}$ of the system and an ancilla. The channels in the second line
are evaluated at $t=t_\ug$. Including the ancilla makes the optimization sensitive to errors in system--ancilla correlations, so
$d_\diamond$ gives a state-independent worst-case channel error. For two quantum channels, $0\leq d_\diamond\leq2$. Diamond-norm
conventions are given in the Supplemental Material~\cite{SupplementalMaterial}.

\paragraph{Driven qubit under correlated noise.---}
We apply the NMQME to a driven qubit subject to longitudinal classical noise. Single-qubit gates are elementary operations,
while transition-frequency fluctuations often contain low-frequency components correlated over a gate
~\cite{Ithier2005,Paladino2014,Nakamura2024}. We consider
\begin{equation}
    \hH_\uq (t) = \frac{1}{2}\left[\omega_0 + \varepsilon \xi (t)\right]\hat{\sigma}_z + f_x (t) \cos(\omega_\ud t) \hat{\sigma}_x,
    \label{eq:HDrivenQubitNoise}
\end{equation}
where $\omega_0$ and $\omega_\ud$ are the qubit and drive frequencies and $f_x(t)$ is the drive envelope. Thus
$\hV=\hat\sigma_z/2$. Although the stochastic term is longitudinal, the control rotates
$\hV_\uI(t)=\hU_0^\dagger(t)\hV\hU_0(t)$ during the gate. The response therefore depends on both the noise correlations and the
control waveform, directly testing their noncommuting interplay.

We consider two zero-mean Gaussian processes with qualitatively different memory. OU noise has an exponentially decaying
correlation function with tunable correlation time $\tau_\uc$~\cite{hanggi95,lehle18}. Band-limited $1/f$ noise instead
distributes slow fluctuations over a broad range of timescales and is relevant to solid-state devices
~\cite{dutta1981,Ithier2005,Paladino2014}. Because the statistical and dynamical components enter separately, the same
Magnus-response construction applies to both models, with their correlation functions supplying the statistical kernel weights.
OU noise probes a single memory time, whereas $1/f$ noise provides a broad, long-lived test of the time-local approximation. The
noise generation and its statistical validation are detailed in the Supplemental Material~\cite{SupplementalMaterial}.

\paragraph{Results.---}
We first identify dimensionless parameters controlling the NMQME expansion by estimating the leading time-integrated noise
contribution; derivations are given in the Supplemental Material~\cite{SupplementalMaterial}. For OU noise,
\begin{equation}
    \delta_\mm{OU} \approx \frac{\sigma^2 \tau_\uc t_\ug}{ \sqrt{1+ \frac{\theta^2}{t_\ug^2} \tau_\uc^2}},
    \label{eq:SmallParameterOU}
\end{equation}
where $\sigma$, $\tau_\uc$, and $t_\ug$ are the noise amplitude, correlation time, and gate duration, while $\theta/t_\ug$ sets
the characteristic control frequency. The two timescales echo the coarse-graining underlying Brownian motion:~$\tau_\uc$
characterizes the persistence of individual fluctuations, while $t_\ug$ determines the interval over which their effect
accumulates. Thus, in the short-memory regime, $\delta_\mm{OU}\sim(\sigma^2\tau_\uc)t_\ug$, the product of an effective
noise-induced rate and the gate duration.

For the band-limited $1/f$ noise considered here,
\begin{equation}
    \delta_{1/f} \approx \frac{\pi}{\theta}\left(\sigma t_\ug\right)^2.
    \label{eq:SmallParameterOneF}
\end{equation}
In contrast to OU noise, the $1/f$ process has no single correlation time from which to define an effective noise-induced rate.
Its low-frequency components remain correlated over the gate and accumulate as slowly varying frequency offsets, giving the
quadratic dependence $\delta_{1/f}\propto(\sigma t_\ug)^2$. The two parameters thus describe distinct memory regimes:
$\delta_\mm{OU}$ has a rate-times-duration form in the short-memory limit, whereas $\delta_{1/f}$ quantifies the accumulated
effect of slow fluctuations that remain correlated throughout the controlled evolution.

Figure~\ref{fig:SemiDefPositiveChoiAll} compares the Choi spectra with stochastic simulations. For OU noise,
Figs.~\ref{fig:SemiDefPositiveChoiAll}(a)--\ref{fig:SemiDefPositiveChoiAll}(d) show that the second- and fourth-order NMQME both
reproduce the reference throughout $\delta_\mm{OU}<1$. The fourth-order minimum eigenvalue becomes negative only at
$\delta_{\mm{OU},\uc}=7.38$, well beyond the expected regime of validity for perturbation theory; the second-order map remains
positive over the displayed range. The same behavior occurs for $1/f$ noise, where fourth-order positivity is lost at
$\delta_{1/f,\uc}=25.7$. The trace-preservation residual vanishes numerically for both models and orders. Thus the NMQME maps are
CPTP and accurate throughout the perturbative regime, although physical admissibility beyond it does not guarantee perturbative
accuracy.

\begin{figure}[t]
    \centering
    \includegraphics[width=0.99\columnwidth]{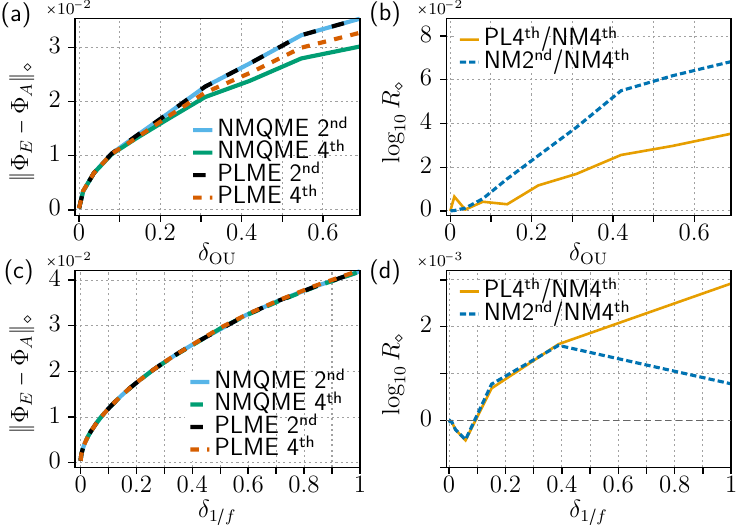}
    \caption{
        Diamond-norm errors relative to reference channels obtained from stochastic simulations for OU [(a),(b)] and band-limited
        $1/f$ noise [(c),(d)]. Panels (a) and (c) compare the second- and fourth-order NMQME and PLME errors. Panels (b) and (d)
        show $\log_{10}(R_\diamond)$ for the indicated error ratios; positive values favor the fourth-order NMQME.
    }
    \label{fig:DiamondNormAll} 
\end{figure}

Figure~\ref{fig:DiamondNormAll} compares the diamond-norm errors of the second- and fourth-order NMQME and PLME channels relative
to the reference channels obtained from stochastic simulations. The sampled noise ensembles underlying these reference channels
are validated against their target covariances and spectra in the Supplemental Material~\cite{SupplementalMaterial}. For OU
noise, the fourth-order NMQME has the smallest error over the perturbative regime, followed by the fourth-order PLME, while the
second-order methods coincide. The positive ratios in Fig.~\ref{fig:DiamondNormAll}(b) show that the relative advantage of the
fourth-order NMQME grows as higher-order corrections become relevant. For band-limited $1/f$ noise, all four errors are nearly
indistinguishable, with only sub-percent differences.  Thus, the two independently organized closure schemes agree quantitatively,
while the functional-response closure gives a modest fourth-order advantage for OU noise. Supplemental average-fidelity and
time-resolved Pauli benchmarks confirm the same perturbative trends, including the accuracy of the Bloch-vector dynamics within
the perturbative regime ~\cite{SupplementalMaterial}.

\paragraph{Conclusion.---}
We solved the perturbative closure problem for driven systems under temporally correlated Gaussian classical noise, obtaining the
NMQME through fourth order. The framework is modular: Donsker--Furutsu--Novikov identities assign statistical weights to
functional-response insertions, while Magnus commutators retain their noncommuting chronological action along the controlled
quantum trajectory. For a driven qubit under OU and band-limited $1/f$ noise, the NMQME reproduces stochastic simulations and
remains CPTP throughout the perturbative regime. It matches the PLME accuracy, with a modest improvement for OU noise. More
generally, connected higher noise cumulants extend the response hierarchy to non-Gaussian classical noise, while a suitable
operator-valued response identity could provide an analogous route to time-local QMEs for quantum environments.

\paragraph{Data availability.---} The code and data required to reproduce the numerical results are available in the
\texttt{QuantumNoiseSimulator} repository on GitHub at~\url{https://github.com/NMQME/NMQME}.

\bibliography{QMEClNoise.bib}

@article{kubo1957,
  author  = {Kubo, Ryogo},
  title   = {Statistical-Mechanical Theory of Irreversible Processes.
             I. General Theory and Simple Applications to Magnetic and
             Conduction Problems},
  journal = {Journal of the Physical Society of Japan},
  volume  = {12},
  number  = {6},
  pages   = {570--586},
  year    = {1957},
  doi     = {10.1143/JPSJ.12.570}
}

@article{martin1973,
  author  = {Martin, P. C. and Siggia, E. D. and Rose, H. A.},
  title   = {Statistical Dynamics of Classical Systems},
  journal = {Physical Review A},
  volume  = {8},
  number  = {1},
  pages   = {423--437},
  year    = {1973},
  doi     = {10.1103/PhysRevA.8.423}
}

@book{heinz07,
  author    = {Breuer, Heinz-Peter and Petruccione, Francesco},
  title     = {The Theory of Open Quantum Systems},
  publisher = {Oxford University Press},
  address   = {Oxford},
  year      = {2007},
  isbn      = {978-0-19-921390-0},
  doi       = {10.1093/acprof:oso/9780199213900.001.0001}
}

@article{Blanes2009,
  author  = {Blanes, Sergio and Casas, Fernando and Oteo, Jos{\'e} A. and Ros, Jos{\'e}},
  title   = {The {Magnus} Expansion and Some of Its Applications},
  journal = {Physics Reports},
  volume  = {470},
  number  = {5--6},
  pages   = {151--238},
  year    = {2009},
  doi     = {10.1016/j.physrep.2008.11.001}
}

@article{lehle18,
  author  = {Lehle, B. and Peinke, J.},
  title   = {Analyzing a Stochastic Process Driven by {Ornstein--Uhlenbeck} Noise},
  journal = {Physical Review E},
  volume  = {97},
  number  = {1},
  pages   = {012113},
  year    = {2018},
  doi     = {10.1103/PhysRevE.97.012113}
}

@article{contreras21,
  author  = {Contreras-Vergara, O. and Lucero-Azuara, N. and S{\'a}nchez-Salas, Norma and Jim{\'e}nez-Aquino, J. I.},
  title   = {Langevin Original Approach and {Ornstein--Uhlenbeck}-Type Processes},
  journal = {Physica A: Statistical Mechanics and its Applications},
  volume  = {584},
  pages   = {126349},
  year    = {2021},
  doi     = {10.1016/j.physa.2021.126349}
}

@incollection{maller09,
  author    = {Maller, Ross A. and M{\"u}ller, Gernot and Szimayer, Alexander},
  title     = {{Ornstein--Uhlenbeck} Processes and Extensions},
  booktitle = {Handbook of Financial Time Series},
  editor    = {Andersen, Torben G. and Davis, Richard A. and Krei{\ss}, Jens-Peter and Mikosch, Thomas},
  publisher = {Springer},
  address   = {Berlin, Heidelberg},
  pages     = {421--437},
  year      = {2009},
  doi       = {10.1007/978-3-540-71297-8_18},
  isbn      = {978-3-540-71296-1}
}

@article{bowdrey02,
  author  = {Bowdrey, Mark D. and Oi, Daniel K. L. and Short, Anthony J. and Banaszek, Konrad and Jones, Jonathan A.},
  title   = {Fidelity of Single-Qubit Maps},
  journal = {Physics Letters A},
  volume  = {294},
  number  = {5--6},
  pages   = {258--260},
  year    = {2002},
  doi     = {10.1016/S0375-9601(02)00069-5}
}

@book{Wilde_2017,
  author    = {Wilde, Mark M.},
  title     = {Quantum Information Theory},
  edition   = {2},
  publisher = {Cambridge University Press},
  address   = {Cambridge},
  year      = {2017},
  doi       = {10.1017/9781316809976}
}

@article{kasdin1995,
  author  = {Kasdin, N. Jeremy},
  title   = {Discrete Simulation of Colored Noise and Stochastic Processes and {$1/f^{\alpha}$} Power Law Noise Generation},
  journal = {Proceedings of the IEEE},
  volume  = {83},
  number  = {5},
  pages   = {802--827},
  year    = {1995},
  doi     = {10.1109/5.381848}
}

@article{dutta1981,
  author  = {Dutta, P. and Horn, P. M.},
  title   = {Low-Frequency Fluctuations in Solids: {$1/f$} Noise},
  journal = {Reviews of Modern Physics},
  volume  = {53},
  number  = {3},
  pages   = {497--516},
  year    = {1981},
  doi     = {10.1103/RevModPhys.53.497}
}

@article{Groszkowski_2023,
  author  = {Groszkowski, Peter and Seif, Alireza and Koch, Jens and Clerk, Aashish A.},
  title   = {Simple Master Equations for Describing Driven Systems Subject to Classical Non-{Markovian} Noise},
  journal = {Quantum},
  volume  = {7},
  pages   = {972},
  year    = {2023},
  doi     = {10.22331/q-2023-04-06-972}
}

@incollection{hanggi95,
  author    = {H{\"a}nggi, Peter and Jung, Peter},
  title     = {Colored Noise in Dynamical Systems},
  booktitle = {Advances in Chemical Physics},
  volume    = {89},
  publisher = {John Wiley \& Sons},
  pages     = {239--326},
  year      = {1995},
  doi       = {10.1002/9780470141489.ch4}
}

@article{Furutsu1963,
  author  = {Furutsu, Koichi},
  title   = {On the Statistical Theory of Electromagnetic Waves in a Fluctuating Medium. Part {I}},
  journal = {Journal of Research of the National Bureau of Standards, Section D: Radio Propagation},
  volume  = {67D},
  number  = {3},
  pages   = {303--323},
  year    = {1963}
}

@article{Novikov1965,
  author  = {Novikov, Evgenii A.},
  title   = {Functionals and the Random-Force Method in Turbulence Theory},
  journal = {Soviet Physics JETP},
  volume  = {20},
  number  = {5},
  pages   = {1290--1294},
  year    = {1965}
}

@incollection{Donsker1964,
    author = {Donsker, Monroe D.},
    title = {On Function Space Integrals},
    booktitle = {Analysis in Function Space: Proceedings of a Conference on the Theory and Applications of Analysis in Function Space},
    editor = {Martin, William Ted and Segal, Irving E.},
    publisher = {MIT Press},
    address = {Cambridge, Massachusetts},
    pages = {17--30},
    year = {1964}
}

@article{Magnus1954,
  author  = {Magnus, Wilhelm},
  title   = {On the Exponential Solution of Differential Equations for a Linear Operator},
  journal = {Communications on Pure and Applied Mathematics},
  volume  = {7},
  number  = {4},
  pages   = {649--673},
  year    = {1954},
  doi     = {10.1002/cpa.3160070404}
}

@article{Jamiolkowski1972,
  author  = {Jamio{\l}kowski, Andrzej},
  title   = {Linear Transformations Which Preserve Trace and Positive
             Semidefiniteness of Operators},
  journal = {Reports on Mathematical Physics},
  volume  = {3},
  number  = {4},
  pages   = {275--278},
  year    = {1972},
  doi     = {10.1016/0034-4877(72)90011-0}
}

@article{Choi1975,
  author  = {Choi, Man-Duen},
  title   = {Completely Positive Linear Maps on Complex Matrices},
  journal = {Linear Algebra and its Applications},
  volume  = {10},
  number  = {3},
  pages   = {285--290},
  year    = {1975},
  doi     = {10.1016/0024-3795(75)90075-0}
}

@article{Ithier2005,
  author  = {Ithier, Gr{\'e}goire and Collin, E. and Joyez, P. and
             Meeson, P. J. and Vion, D. and Esteve, D. and Chiarello, F.
             and Shnirman, A. and Makhlin, Y. and Schriefl, J. and
             Sch{\"o}n, G.},
  title   = {Decoherence in a Superconducting Quantum Bit Circuit},
  journal = {Physical Review B},
  volume  = {72},
  number  = {13},
  pages   = {134519},
  year    = {2005},
  doi     = {10.1103/PhysRevB.72.134519}
}

@article{Paladino2014,
  author  = {Paladino, Elisabetta and Galperin, Yuri M. and Falci, Giuseppe
             and Altshuler, Boris L.},
  title   = {{$1/f$} Noise: Implications for Solid-State Quantum Information},
  journal = {Reviews of Modern Physics},
  volume  = {86},
  number  = {2},
  pages   = {361--418},
  year    = {2014},
  doi     = {10.1103/RevModPhys.86.361}
}

@inproceedings{Aharonov1998,
  author    = {Aharonov, Dorit and Kitaev, Alexei and Nisan, Noam},
  title     = {Quantum Circuits with Mixed States},
  booktitle = {Proceedings of the Thirtieth Annual ACM Symposium on
               Theory of Computing},
  pages     = {20--30},
  publisher = {Association for Computing Machinery},
  address   = {New York},
  year      = {1998},
  doi       = {10.1145/276698.276708}
}

@book{Watrous2018,
  author    = {Watrous, John},
  title     = {The Theory of Quantum Information},
  publisher = {Cambridge University Press},
  address   = {Cambridge},
  year      = {2018},
  isbn      = {978-1-107-18056-7}
}

@article{PuzzuoliWatrous2017,
  author  = {Puzzuoli, Daniel and Watrous, John},
  title   = {Ancilla Dimension in Quantum Channel Discrimination},
  journal = {Annales Henri Poincar{\'e}},
  volume  = {18},
  number  = {4},
  pages   = {1153--1184},
  year    = {2017},
  doi     = {10.1007/s00023-016-0537-y}
}

@article{Gorini1976,
  author  = {Gorini, Vittorio and Kossakowski, Andrzej and
             Sudarshan, E. C. G.},
  title   = {Completely Positive Dynamical Semigroups of
             {$N$}-Level Systems},
  journal = {Journal of Mathematical Physics},
  volume  = {17},
  number  = {5},
  pages   = {821--825},
  year    = {1976},
  doi     = {10.1063/1.522979}
}

@article{Lindblad1976,
  author  = {Lindblad, G{\"o}ran},
  title   = {On the Generators of Quantum Dynamical Semigroups},
  journal = {Communications in Mathematical Physics},
  volume  = {48},
  number  = {2},
  pages   = {119--130},
  year    = {1976},
  doi     = {10.1007/BF01608499}
}

@article{Nakajima1958,
  author  = {Nakajima, Sadao},
  title   = {On Quantum Theory of Transport Phenomena:
             Steady Diffusion},
  journal = {Progress of Theoretical Physics},
  volume  = {20},
  number  = {6},
  pages   = {948--959},
  year    = {1958},
  doi     = {10.1143/PTP.20.948}
}

@article{Zwanzig1960,
  author  = {Zwanzig, Robert},
  title   = {Ensemble Method in the Theory of Irreversibility},
  journal = {Journal of Chemical Physics},
  volume  = {33},
  number  = {5},
  pages   = {1338--1341},
  year    = {1960},
  doi     = {10.1063/1.1731409}
}

@article{Shibata1977,
  author  = {Shibata, Fumio and Takahashi, Yosuke and
             Hashitsume, Natsuki},
  title   = {A Generalized Stochastic Liouville Equation.
             Non-{Markovian} versus Memoryless Master Equations},
  journal = {Journal of Statistical Physics},
  volume  = {17},
  number  = {4},
  pages   = {171--187},
  year    = {1977},
  doi     = {10.1007/BF01040100}
}

@article{Chaturvedi1979,
  author  = {Chaturvedi, S. and Shibata, F.},
  title   = {Time-Convolutionless Projection Operator Formalism for
             Elimination of Fast Variables. Applications to
             {Brownian} Motion},
  journal = {Zeitschrift f{\"u}r Physik B},
  volume  = {35},
  number  = {3},
  pages   = {297--308},
  year    = {1979},
  doi     = {10.1007/BF01319852}
}

@article{Hangleiter2021,
  author  = {Hangleiter, Tobias and Cerfontaine, Pascal and
             Bluhm, Hendrik},
  title   = {Filter-Function Formalism and Software Package to
             Compute Quantum Processes of Gate Sequences for
             Classical Non-{Markovian} Noise},
  journal = {Physical Review Research},
  volume  = {3},
  number  = {4},
  pages   = {043047},
  year    = {2021},
  doi     = {10.1103/PhysRevResearch.3.043047}
}

@article{Chalermpusitarak2021,
  author  = {Chalermpusitarak, Teerawat and Tonekaboni, Behnam and
             Wang, Yuanlong and Norris, Leigh M. and Viola, Lorenza
             and Paz-Silva, Gerardo A.},
  title   = {Frame-Based Filter-Function Formalism for Quantum
             Characterization and Control},
  journal = {PRX Quantum},
  volume  = {2},
  number  = {3},
  pages   = {030315},
  year    = {2021},
  doi     = {10.1103/PRXQuantum.2.030315}
}

@article{Nakamura2024,
  author  = {Nakamura, Kiyoto and Ankerhold, Joachim},
  title   = {Gate Operations for Superconducting Qubits and
             Non-{Markovianity}},
  journal = {Physical Review Research},
  volume  = {6},
  number  = {3},
  pages   = {033215},
  year    = {2024},
  doi     = {10.1103/PhysRevResearch.6.033215}
}

@article{AthanassoulisMamis2019,
  author  = {Athanassoulis, Gerassimos A. and Mamis, Konstantinos I.},
  title   = {Extensions of the {Novikov--Furutsu} Theorem, Obtained by
             Using {Volterra} Functional Calculus},
  journal = {Physica Scripta},
  volume  = {94},
  number  = {11},
  pages   = {115217},
  year    = {2019},
  doi     = {10.1088/1402-4896/ab10b5}
}

@article{nv-gillespie1996,
  author  = {Gillespie, Daniel T.},
  title   = {Exact numerical simulation of the {O}rnstein-{U}hlenbeck process and its integral},
  journal = {Phys. Rev. E},
  volume  = {54},
  pages   = {2084--2091},
  year    = {1996},
  doi     = {10.1103/PhysRevE.54.2084}
}

@article{nv-hosking1981,
  author  = {Hosking, J. R. M.},
  title   = {Fractional differencing},
  journal = {Biometrika},
  volume  = {68},
  number  = {1},
  pages   = {165--176},
  year    = {1981},
  doi     = {10.1093/biomet/68.1.165}
}

@article{nv-timmer1995,
  author  = {Timmer, J. and K{\"o}nig, M.},
  title   = {On generating power law noise},
  journal = {Astron. Astrophys.},
  volume  = {300},
  pages   = {707--710},
  year    = {1995}
}

@article{nv-shinozuka1991,
  author  = {Shinozuka, Masanobu and Deodatis, George},
  title   = {Simulation of Stochastic Processes by Spectral Representation},
  journal = {Appl. Mech. Rev.},
  volume  = {44},
  number  = {4},
  pages   = {191--204},
  year    = {1991},
  doi     = {10.1115/1.3119501}
}

@article{nv-dunn1961,
  author  = {Dunn, Olive Jean},
  title   = {Multiple Comparisons among Means},
  journal = {J. Am. Stat. Assoc.},
  volume  = {56},
  number  = {293},
  pages   = {52--64},
  year    = {1961},
  doi     = {10.1080/01621459.1961.10482090}
}

@article{nv-blackman2021,
  author  = {Blackman, David and Vigna, Sebastiano},
  title   = {Scrambled Linear Pseudorandom Number Generators},
  journal = {ACM Trans. Math. Softw.},
  volume  = {47},
  number  = {4},
  pages   = {36:1--36:32},
  year    = {2021},
  doi     = {10.1145/3460772}
}

@article{nv-johansson2013,
  author  = {Johansson, J. R. and Nation, P. D. and Nori, Franco},
  title   = {{QuTiP} 2: A {P}ython framework for the dynamics of open quantum systems},
  journal = {Comput. Phys. Commun.},
  volume  = {184},
  number  = {4},
  pages   = {1234--1240},
  year    = {2013},
  doi     = {10.1016/j.cpc.2012.11.019}
}

@book{papoulis2002,
  title={Probability, Random Variables, and Stochastic Processes},
  author={Papoulis, Athanasios and Pillai, S. Unnikrishna},
  edition={4th},
  year={2002},
  publisher={McGraw-Hill},
  address={Boston},
  isbn={978-0073660110}
}

@misc{SupplementalMaterial,
  note = {See Supplemental Material for the perturbative derivation of
  the NMQME, channel-diagnostic conventions, derivations of the
  expansion parameters, additional average-fidelity and time-resolved
  Bloch-vector results, and the numerical generation and validation of
  the OU and band-limited $1/f$ noise ensembles.}
}

@article{Ishizaki2009,
  author  = {Ishizaki, Akihito and Fleming, Graham R.},
  title   = {Unified Treatment of Quantum Coherent and Incoherent
             Hopping Dynamics in Electronic Energy Transfer:
             Reduced Hierarchy Equation Approach},
  journal = {J. Chem. Phys.},
  volume  = {130},
  pages   = {234111},
  year    = {2009},
  doi     = {10.1063/1.3155372}
}

\clearpage
\onecolumngrid
\section*{Supplemental Material for:~``Novikov-Magnus Master Equation for Driven Quantum Systems under Classical Correlated Noise''}
\setcounter{secnumdepth}{2}
\setcounter{section}{0}
\setcounter{equation}{0}
\setcounter{figure}{0}
\setcounter{table}{0}

\renewcommand{\thesection}{S\arabic{section}}
\renewcommand{\theequation}{S\arabic{equation}}
\renewcommand{\thefigure}{S\arabic{figure}}
\renewcommand{\thetable}{S\arabic{table}}

\newcommand{\numSeed}{20260926}   
\newcommand{\numOuSigma}{1}   
\newcommand{\numOuTauC}{1}   
\newcommand{\numOuDt}{0.01}   
\newcommand{\numOuM}{1000}   
\newcommand{\numOuFCorner}{0.1592}   
\newcommand{\numBlSigma}{1}   
\newcommand{\numBlDt}{0.005}   
\newcommand{\numBlFl}{0.1592}   
\newcommand{\numBlFh}{15.92}   
\newcommand{\numBlVarExact}{9.21}   
\newcommand{\numBlM}{1000}   
\newcommand{\numPinkAlpha}{1}   
\newcommand{\numPinkQ}{0.001}   
\newcommand{\numPinkDt}{0.001}   
\newcommand{\numPinkM}{1000}   
\newcommand{\numOuN}{2001}   
\newcommand{\numOuT}{20}   
\newcommand{\numBlN}{2001}   
\newcommand{\numBlT}{10}   
\newcommand{\numBlL}{16200}   
\newcommand{\numBlNbins}{1277}   
\newcommand{\numPinkN}{4001}   
\newcommand{\numPinkT}{4}   
\newcommand{\numOuKsD}{0.02865}   
\newcommand{\numOuKsCrit}{0.04294}   
\newcommand{\numOuSkew}{0.1091}   
\newcommand{\numOuSkewSe}{0.07746}   
\newcommand{\numOuKurt}{-0.3185}   
\newcommand{\numOuKurtSe}{0.1549}   
\newcommand{\numOuTauCHat}{0.9931}   
\newcommand{\numOuTauCHatSe}{0.002294}   
\newcommand{\numOuFitMaxlag}{2.9}   
\newcommand{\numOuAcfMaxz}{1.986}   
\newcommand{\numOuPsdMaxz}{3.462}   
\newcommand{\numBlTLong}{200}   
\newcommand{\numBlTZoom}{0.4}   
\newcommand{\numBlSd}{3.035}   
\newcommand{\numBlVarTEmp}{8.904}   
\newcommand{\numBlAcfMaxz}{1.845}   
\newcommand{\numBlCiMaxdiff}{\ensuremath{1.81\times10^{-4}}}   
\newcommand{\numBlTwotimeMaxz}{3.787}   
\newcommand{\numBlTwotimeZlim}{5.162}   
\newcommand{\numBlPsdMaxz}{2.813}   
\newcommand{\numBlBetaHat}{0.9945}   
\newcommand{\numBlBetaSe}{0.006956}   
\newcommand{\numBlBetaTarget}{1.001}   
\newcommand{\numBlFitFmin}{0.6997}   
\newcommand{\numBlFitFmax}{7.896}   
\newcommand{\numPinkVarTExact}{0.003706}   
\newcommand{\numPinkVarTEmp}{0.003595}   
\newcommand{\numPinkVarZeroExact}{0.001}   
\newcommand{\numPinkTwotimeMaxz}{3.632}   
\newcommand{\numPinkTwotimeZlim}{5.162}   
\newcommand{\numPinkLowbinExcess}{1.101}   
\newcommand{\numPinkPsdMaxz}{3.432}   
\newcommand{\numPinkAcfMaxz}{0.8549}   
\newcommand{\numConvPool}{6000}   
\newcommand{\numConvRmax}{60}   
\newcommand{\numOuValid}{1}   
\newcommand{\numOuZMean}{1.022}   
\newcommand{\numOuZlimMean}{3.291}   
\newcommand{\numOuZVar}{0.8634}   
\newcommand{\numOuZlimVar}{3.291}   
\newcommand{\numOuZAcov}{1.986}   
\newcommand{\numOuZlimAcov}{4.79}   
\newcommand{\numOuZPsd}{3.462}   
\newcommand{\numOuZlimPsd}{4.892}   
\newcommand{\numBlValid}{1}   
\newcommand{\numBlZMean}{0.6637}   
\newcommand{\numBlZlimMean}{3.291}   
\newcommand{\numBlZVar}{0.7428}   
\newcommand{\numBlZlimVar}{3.291}   
\newcommand{\numBlZAcov}{1.845}   
\newcommand{\numBlZlimAcov}{4.79}   
\newcommand{\numBlZPsd}{2.813}   
\newcommand{\numBlZlimPsd}{4.892}   
\newcommand{\numPinkValid}{1}   
\newcommand{\numPinkZMean}{0.02685}   
\newcommand{\numPinkZlimMean}{3.291}   
\newcommand{\numPinkZVar}{0.6689}   
\newcommand{\numPinkZlimVar}{3.291}   
\newcommand{\numPinkZAcov}{0.8549}   
\newcommand{\numPinkZlimAcov}{4.928}   
\newcommand{\numPinkZPsd}{3.432}   
\newcommand{\numPinkZlimPsd}{5.026}   

\providecommand{\nvVar}{\mathrm{Var}}
\providecommand{\nvCov}{\mathrm{Cov}}
\providecommand{\nvCi}{\mathrm{Ci}}

\section{Perturbative derivation of the NMQME}
\label{sec:supp-fullEoM}

This section provides the technical steps leading from the exact Magnus-response identity in Eq.~\eqref{eq:FunDerLIMagnusFinal} to
the second- and fourth-order contributions to the time-local NMQME in Eq.~\eqref{eq:finalEOMKernel}. We first expand the
functional response to the required perturbative order and then apply the Donsker--Furutsu--Novikov relation recursively to
eliminate the remaining explicit noise factors. We illustrate this procedure with a representative contribution involving
$\Omega_2[\xi(t)]$ and explain how the resulting partially ordered integration domains are resolved into compatible chronological
sectors. The same steps apply to the remaining terms. To keep the presentation tractable, we retain the intermediate structure
needed to establish the closure and the complete fourth-order equation of motion, but omit algebraic manipulations that follow by
direct repetition of the illustrated procedure.

As in the main text, we write the stochastic interaction-picture map in Magnus form,
\begin{equation} 
    \sL_\uI[\xi(t)] = \exp\left\{\Omega[\xi(t)]\right\}, 
    \label{eq:supp-LIMagnus} 
\end{equation} 
The Magnus exponent is expanded perturbatively as $\Omega[\xi(t)] =\sum_{n=1}^{\infty}\varepsilon^n\Omega_n[\xi(t)]$. The first
two terms of the expansion are~\cite{Blanes2009}
\begin{equation}
	\begin{aligned}
        \Omega_1 [\xi(t)] &= \int_0^t  dt_1 \, \xi(t_1) \, \slr_\uI (t_1) ,\\
		\Omega_2 [\xi (t)] &= \frac{1}{2} \int_0^t dt_1 \int_0^{t_1} dt_2 \, \xi (t_1)  \,  \xi (t_2) \left[ \slr_\uI (t_1), \slr_\uI (t_2)  \right]  .
	\end{aligned}
	\label{eq:MagnusExpansion}
\end{equation}
The ordered integration domain in the second line of Eq.~\eqref{eq:MagnusExpansion}, together with the nested commutators
appearing at higher orders, retains the chronological structure generated by the generally noncommuting superoperators
$\slr_\uI(t)$.

To evaluate the functional derivative entering Eq.~\eqref{eq:NovikovZeroMeanGaussian}, we express the operator exponential as the
finite-product limit 
\begin{equation} 
    \sL_\uI[\xi(t)] = \lim_{n\to\infty} \left[ \exp\left\{ \frac{\Omega[\xi(t)]}{n} \right\} \right]^n. 
    \label{eq:FuncDerLIMagnus} 
\end{equation}
Applying the product rule for functional derivatives gives
\begin{equation} 
    \begin{aligned} 
        \frac{\delta\sL_\uI[\xi(t)]}{\delta\xi(t_1)} = \lim_{n\to\infty} \sum_{l=0}^{n-1} 
        \left[\exp\left(\frac{\Omega[\xi(t)]}{n}\right) \right]^{n-l-1}
        \left[\frac{\delta}{\delta\xi(t_1)}\exp\left(\frac{\Omega[\xi(t)]}{n}\right)\right] 
        \left[ \exp\left(\frac{\Omega[\xi(t)]}{n}\right) \right]^l. 
    \end{aligned} 
    \label{eq:FuncDerLIMagnusProduct} 
\end{equation}
Using $\mm{Ad}_{e^{\hO}} = e^{\ad_{\hO}}$ to move the differentiated exponential through the preceding factors, and expanding that
exponential in powers of $\Omega[\xi(t)]/n$, we obtain
\begin{equation} 
    \begin{aligned} 
        \frac{\delta\sL_\uI[\xi(t)]}{\delta\xi(t_1)} &= \left\{ \lim_{n\to\infty} \frac{1}{n} \sum_{l=0}^{n-1} \left[
        \exp\left(\frac{\ad_{\Omega[\xi(t)]}}{n} \right) \right]^l \right\} 
        \left[\frac{\delta}{\delta\xi(t_1)} \left( \lim_{n\to\infty} \sum_{k=1}^{\infty} \frac{1}{k!} \frac{
        \left(\Omega[\xi(t)]\right)^k }{ n^{k-1} } \right)\right]
        \lim_{n\to\infty} \left[ \exp\left( \frac{\Omega[\xi(t)]}{n} \right) \right]^{n-1}. 
    \end{aligned} 
    \label{eq:FuncDerLIMagnus4} 
\end{equation}
For the finite-dimensional operators and superoperators considered here, the three limits exist in the operator norm. Operator
multiplication and superoperator action are continuous in this norm, so the limit may be evaluated factor by factor.

The first factor contains the noncommuting operator structure. Expanding the adjoint exponential gives
\begin{equation} 
    \begin{aligned} 
        \lim_{n\to\infty} \frac{1}{n} \sum_{l=0}^{n-1} \left[ \exp\left(\frac{\ad_{\Omega[\xi(t)]}}{n}\right) \right]^l 
        &= \sum_{k=0}^{\infty} \frac{\ad_{\Omega[\xi(t)]}^k}{k!} \lim_{n\to\infty} \frac{1}{n} \sum_{l=0}^{n-1} \left(\frac{l}{n}\right)^k \\ 
        &= \sum_{k=0}^{\infty} \frac{\ad_{\Omega[\xi(t)]}^k}{(k+1)!}. 
    \end{aligned} 
    \label{eq:supp-AdjointGeometricLimit} 
\end{equation}
In the second equality, the normalized sum becomes the Riemann integral $\int_0^1\di{u}\,u^k=1/(k+1)$. The resulting nested
adjoint actions retain the noncommuting chronological structure of the quantum response.

In the second factor of Eq.~\eqref{eq:FuncDerLIMagnus4}, only the $k=1$ term survives as $n\to\infty$, because every term with
$k \geq 2$ contains a positive inverse power of $n$. Thus, this factor becomes $\delta\Omega[\xi(t)]/\delta\xi(t_1)$. The third
factor recovers the stochastic map,
\begin{equation} 
    \lim_{n\to\infty} \left[ \exp\left( \frac{\Omega[\xi(t)]}{n} \right) \right]^{n-1} = \exp\left(\Omega[\xi(t)]\right) = \sL_\uI[\xi(t)]. 
    \label{eq:supp-MapFactorLimit} 
\end{equation} 
Combining the three factors yields
\begin{equation} 
    \frac{\delta\sL_\uI[\xi(t)]}{\delta\xi(t_1)} = \sum_{k=0}^{\infty} \frac{1}{(k+1)!} \ad_{\Omega[\xi(t)]}^{k} \left[ \frac{\delta\Omega[\xi(t)]}{\delta\xi(t_1)} \right] \sL_\uI[\xi(t)]. 
    \label{eq:supp-FunDerLIMagnusFinal} 
\end{equation}
This is the exact Magnus-response identity stated in Eq.~\eqref{eq:FunDerLIMagnusFinal} of the main text. We use
$\ad_{\hA}^0(\hB)=\hB$ and define the higher nested commutators recursively by $\ad_{\hA}^k(\hB)=\com{\hA}{\ad_{\hA}^{k-1}(\hB)}$.

Substituting the perturbative expansion of the Magnus exponent into Eq.~\eqref{eq:supp-FunDerLIMagnusFinal} and collecting equal
powers of $\varepsilon$ gives
\begin{equation} 
    \begin{aligned} 
        \frac{\delta\sL_\uI[\xi(t)]}{\delta\xi(t_1)} &= \left\{ \varepsilon \frac{\delta\Omega_1[\xi(t)]}{\delta\xi(t_1)} \right.
        + \varepsilon^2 \left(\frac{\delta\Omega_2[\xi(t)]}{\delta\xi(t_1)} + \frac{1}{2} \com{\Omega_1[\xi(t)]}{\frac{\delta\Omega_1[\xi(t)]} {\delta\xi(t_1)}} \right)
        + \varepsilon^3 \left( \frac{\delta\Omega_3[\xi(t)]}{\delta\xi(t_1)} + \frac{1}{2}
        \com{\Omega_1[\xi(t)]}{\frac{\delta\Omega_2[\xi(t)]} {\delta\xi(t_1)}} \right. \\
        &\phantom{={}}
        \left. \left.
        + \frac{1}{2} \com{\Omega_2[\xi(t)]}{\frac{\delta\Omega_1[\xi(t)]}{\delta\xi(t_1)}}
        + \frac{1}{6} \com{\Omega_1[\xi(t)]} { \com{\Omega_1[\xi(t)]} { \frac{\delta\Omega_1[\xi(t)]} {\delta\xi(t_1)} } } 
        \right)
        \right\} \sL_\uI[\xi(t)] + \BigO\left(\varepsilon^4\right). 
    \end{aligned} 
    \label{eq:supp-FunctionalResponseExpansion} 
\end{equation} 
Equation~\eqref{eq:supp-FunctionalResponseExpansion} contains the functional response through third order in $\varepsilon$.
Because the averaged equation of motion in Eq.~\eqref{eq:AvgLI} contains an additional factor of $\varepsilon$, the retained
response terms generate the NMQME through fourth order in the noise strength.

Inserting Eq.~\eqref{eq:supp-FunctionalResponseExpansion} into Eq.~\eqref{eq:NovikovZeroMeanGaussian} gives the equation of motion
before the remaining mixed averages are resolved,
\begin{equation} 
    \begin{aligned} 
        \partial_t\avg{\sL_\uI(t)}_\xi &= \varepsilon\slr_\uI(t) \int_0^t\di{t_1} \avg{\xi (t) \xi(t_1)}_\xi \left\langle \left\{\varepsilon \frac{\delta\Omega_1[\xi(t)]}{\delta\xi(t_1)}
        + \varepsilon^2 \left(\frac{\delta\Omega_2[\xi(t)]}{\delta\xi(t_1)} + \frac{1}{2} \com{\Omega_1[\xi(t)]}{\frac{\delta\Omega_1[\xi(t)]}{\delta\xi(t_1)}} \right) \right.\right. \\
        &\phantom{={} \varepsilon\slr_\uI(t) \int_0^t\di{t_1}}
        + \varepsilon^3 \left( \frac{\delta\Omega_3[\xi(t)]}{\delta\xi(t_1)} + \frac{1}{2} \com{\Omega_1[\xi(t)]} { \frac{\delta\Omega_2[\xi(t)]} {\delta\xi(t_1)} } 
        + \frac{1}{2} \com{\Omega_2[\xi(t)]} { \frac{\delta\Omega_1[\xi(t)]} {\delta\xi(t_1)} } \right. \\
        &\phantom{={} \varepsilon\slr_\uI(t) \int_0^t\di{t_1}}
        \left. \left. \left.
            + \frac{1}{6} \com{\Omega_1[\xi(t)]} { \com{\Omega_1[\xi(t)]} { \frac{\delta\Omega_1[\xi(t)]} {\delta\xi(t_1)} } } \right)
        \right\} \sL_\uI[\xi(t)] \right\rangle_\xi + \BigO\left(\varepsilon^5\right). 
    \end{aligned} 
    \label{eq:EoMGeneric} 
\end{equation}
Equation~\eqref{eq:EoMGeneric} is not yet closed because the Magnus terms and the stochastic map inside the ensemble average
remain functionals of the noise history. Explicit noise factors are removed by repeatedly applying the Donsker--Furutsu--Novikov
relation to the remaining mixed averages. Each application replaces an explicit noise factor by a correlation function and an
additional functional response. The next subsection illustrates this recursive step for the contribution involving
$\Omega_2[\xi(t)]$.

\subsection{Example:~unravelling the average involving $\Omega_2[\xi(t)]$}

We illustrate the recursive closure by evaluating one of the mixed averages appearing in Eq.~\eqref{eq:EoMGeneric}. From the
second line of Eq.~\eqref{eq:MagnusExpansion}, the functional derivative of $\Omega_2[\xi(t)]$ is
\begin{equation}
    \frac{\delta \Omega_2[\xi(t)]}{\delta\xi(t_1)} = 
    \frac{1}{2} \int_0^t \di{\tau_1} \int_0^{\tau_1} \di{\tau_2} \frac{\delta \xi(\tau_1)}{\delta \xi(t_1)}\xi(\tau_2) [\slr_\uI (\tau_1), \slr_\uI (\tau_2)] 
    + \frac{1}{2} \int_0^t \di{\tau_1} \int_0^{\tau_1} \di{\tau_2} \xi(\tau_1)\frac{\delta\xi(\tau_2)}{\delta \xi(t_1)}[\slr_\uI(\tau_1), \slr_\uI (\tau_2)].
    \label{eq:Y20fctder}
\end{equation}
Using $\delta\xi(\tau)/\delta\xi(t_1)=\delta(\tau-t_1)$ and carrying out the corresponding integrations gives
\begin{equation}
    \frac{\delta \Omega_2[\xi(t)]}{\delta\xi(t_1)} = 
    \frac{1}{2} \int_0^{t_1} \di{\tau_2} \xi(\tau_2) [\slr_\uI (t_1), \slr_\uI (\tau_2)] \Theta (t-t_1)
    + \frac{1}{2} \int_0^t \di{\tau_1} \xi(\tau_1) [\slr_\uI (\tau_1), \slr_\uI (t_1)] \Theta (\tau_1-t_1).
    \label{eq:omega2Example}
\end{equation}
The Heaviside functions make the causal support of the two terms explicit. In the first term, the insertion at $t_1$ must lie
within the evolution interval, while the remaining noise time satisfies $\tau_2\leq t_1$. In the second term, the remaining noise
time satisfies $\tau_1\geq t_1$.

Multiplying Eq.~\eqref{eq:omega2Example} by the stochastic map and averaging over noise realizations gives
\begin{equation}
    \begin{aligned}
        \left\langle \frac{\delta \Omega_2[\xi(t)]}{\delta\xi(t_1)} \sL_\uI (t) \right\rangle_\xi &=
        \frac{1}{2} \int_0^{t_1} \di{\tau_2} [\slr_\uI (t_1), \slr_\uI (\tau_2)] \Theta (t-t_1) \langle\xi(\tau_2)\sL_\uI
        (t)\rangle_\xi \\
        &\phantom{={}}
        + \frac{1}{2} \int_0^t \di{\tau_1} [\slr_\uI (\tau_1), \slr_\uI (t_1)] \Theta (\tau_1-t_1) \langle\xi(\tau_1)\sL_\uI (t)\rangle_\xi,
    \end{aligned}
\end{equation}
We evaluate the remaining noise--map correlations by applying the Donsker--Furutsu--Novikov relation again [see
Eq.~\eqref{eq:NovikovZeroMeanGaussian} of the main text]. In Eq.~\eqref{eq:EoMGeneric}, this mixed average is multiplied by
$\varepsilon^2$, in addition to the overall factor of $\varepsilon$ in the equation of motion. To obtain the fourth-order
contribution, it is therefore sufficient to retain the leading, order-$\varepsilon$ term of the functional response in
Eq.~\eqref{eq:supp-FunDerLIMagnusFinal}. We then find
\begin{equation}
    \begin{aligned}
        \left\langle \frac{\delta \Omega_2[\xi(t)]}{\delta\xi(t_1)} \sL_\uI (t) \right\rangle_\xi & = 
        \varepsilon \left\{\frac{1}{2} \int_0^{t_1} \di{\tau_2} [\slr_\uI (t_1), \slr_\uI (\tau_2)] \Theta(t-t_1)
        \int_0^t \di{t_2} \langle \xi(\tau_2) \xi(t_2) \rangle_\xi \slr_\uI (t_2)  \right. \\
        &\phantom{={}} 
        \left. + \frac{1}{2} \int_0^t \di{\tau_1} [\slr_\uI (\tau_1), \slr_\uI (t_1)] \Theta(\tau_1-t_1)
        \int_0^t \di{t_2} \langle \xi(\tau_1) \xi(t_2) \rangle_\xi \slr_\uI (t_2)\right\} \\
        &\phantom{={}} 
        \times \Theta(t-t_2) \langle \sL_\uI (t)\rangle_\xi.
    \end{aligned}
    \label{eq:Y20Lavg01}
\end{equation}
Equation~\eqref{eq:Y20Lavg01} illustrates the recursive closure. The second application of the Donsker--Furutsu--Novikov relation
replaces the remaining explicit noise factor by a second correlation function and introduces the response insertion
$\slr_\uI(t_2)$. The explicit factor of $\varepsilon$ is the leading order of this additional functional response. Together with
the factor $\varepsilon^2$ multiplying this mixed average and the overall factor of $\varepsilon$ in Eq.~\eqref{eq:EoMGeneric}, it
produces an $\varepsilon^4$ contribution.

The time $t_2$ is integrated independently over the causal interval $0\leq t_2\leq t$. Its position relative to the times already
ordered within $\Omega_2[\xi(t)]$ is therefore not fixed. Since the corresponding superoperators generally do not commute, this
relative ordering must be resolved explicitly.

\subsection{Chronological resolution of the integration domains}

The Magnus expansion fixes the chronological order of the time variables contained within each Magnus term. Recursive applications
of the Donsker--Furutsu--Novikov relation, however, introduce additional response times that are integrated independently. The
combined integration domain is therefore only partially ordered.

The Heaviside functions appearing in the response make its causal structure explicit by restricting each noise variation to times
that can influence the subsequent evolution. This retarded functional-response structure is standard in statistical response
theory~\cite{kubo1957,martin1973}. Causality alone, however, does not fix the relative order of all the independently integrated
response times. Because the associated quantum actions generally do not commute, the partially ordered domain must be decomposed
into all chronological sectors compatible with the constraints already imposed by the Magnus integrals~\cite{Blanes2009}.

The simplest example involves two independently integrated times. The time-ordering operator resolves their square integration
domain into the two chronological sectors
\begin{equation}
    \begin{aligned}
        \hT\left[\int_0^{t} \di{\tau_1} \int_0^{t} \di{\tau_2} \slr_\uI (\tau_1)\slr_\uI (\tau_2)\right]
        & \coloneq \int_0^{t} \di{\tau_1} \int_0^{t} \di{\tau_2} \Theta(\tau_1 - \tau_2) \slr_\uI(\tau_1)\slr_\uI (\tau_2) \\
        &\phantom{={}}
        + \int_0^{t} \di{\tau_1}\int_0^{t} \di{\tau_2} \Theta(\tau_2 - \tau_1) \slr_\uI(\tau_2)\slr_\uI (\tau_1) \\
        &= \int_0^{t} \di{\tau_1} \int_0^{\tau_1} \di{\tau_2} \slr_\uI(\tau_1)\slr_\uI (\tau_2) + \int_0^{t} \di{\tau_2}\int_0^{\tau_2} \di{\tau_1} \slr_\uI(\tau_2)\slr_\uI (\tau_1)
    \label{eq:TimeOrderingHeaviside}
    \end{aligned}
\end{equation}
The two Heaviside functions select the sectors $\tau_2\leq\tau_1$ and $\tau_1\leq\tau_2$, respectively. Apart from their common
boundary, these sectors are disjoint and together cover the full square integration domain. The superoperator product is ordered
differently in the two sectors because exchanging the time arguments also exchanges the order of the generally noncommuting
quantum actions.

The fourth-order construction requires the corresponding resolution of partially ordered three-time domains. For example, the
first term in Eq.~\eqref{eq:Y20Lavg01} already imposes $\tau_2\leq t_1$ through the Magnus domain, while the recursively
introduced time $t_2$ is independently integrated over $[0,t]$. The combined domain is
\begin{equation} 
    \int_0^t\di{t_1} \int_0^{t_1}\di{\tau_2} \int_0^t\di{t_2}. 
    \label{eq:supp-PartiallyOrderedDomain} 
\end{equation}
The position of $t_2$ relative to $\tau_2$ and $t_1$ is unspecified. The domain must therefore be resolved into the three sectors
compatible with the existing constraint $\tau_2\leq t_1$:
\begin{equation} 
    \int_0^t\di{t_1} \int_0^{t_1}\di{\tau_2} \int_0^t\di{t_2} =
    \int_0^t\di{t_1} \int_0^{t_1}\di{\tau_2} \int_0^{\tau_2}\di{t_2}
    + \int_0^t\di{t_1} \int_0^{t_1}\di{t_2} \int_0^{t_2}\di{\tau_2} 
    + \int_0^t\di{t_2} \int_0^{t_2}\di{t_1} \int_0^{t_1}\di{\tau_2}.
    \label{eq:supp-ThreeChronologicalSectors} 
\end{equation}
The three terms correspond, respectively, to $t_2\leq\tau_2\leq t_1$, $\tau_2\leq t_2\leq t_1$, and $\tau_2\leq t_1\leq t_2$.
Within each sector, the response superoperators are placed in the corresponding chronological order. This procedure matches each
classical insertion history to the correctly ordered quantum response, as illustrated in Fig.~\ref{fig:SystemBath}(b) of the main
text.

The second term in Eq.~\eqref{eq:Y20Lavg01} is treated in the same way. There, the Magnus contribution imposes $t_1\leq\tau_1$,
while the recursively introduced time $t_2$ is independent. Its integration domain is therefore resolved into the three sectors
compatible with $t_1\leq\tau_1$. Terms containing three initially independent response times require all six chronological
sectors. More generally, each partially ordered domain is decomposed into every chronological sector consistent with the
constraints inherited from the Magnus integrals.

We apply this chronological decomposition to all contributions in Eq.~\eqref{eq:EoMGeneric}. After the remaining noise factors
have been eliminated recursively, the explicit equation of motion through fourth order is
\begin{equation}
    \begin{aligned}
        \partial_t \avg{\sL_\uI (t)}_\xi
        &= \slr_\uI (t) \left\{ \varepsilon^2 \int_{0}^{t} \di{t_1} \avg{\xi(t) \xi(t_1)}_\xi \slr_\uI (t_1) \right. \\
        &\phantom{={}}
        +\frac{\varepsilon^4}{2} \left( \int_{0}^{t} \di{t_1} \int_0^{t_1} \di{\tau_2} \int_0^{\tau_2} \di{t_2} + \int_{0}^{t}
        \di{t_1} \int_0^{t_1} \di{t_2} \int_0^{t_2} \di{\tau_2} + \int_{0}^{t} \di{t_2} \int_0^{t_2} \di{t_1} \int_0^{t_1}
        \di{\tau_2} \right) \\
        &\phantom{= + \varepsilon^4/2} 
        \times \avg{\xi(t)\xi(t_1)}_\xi \avg{\xi(\tau_2) \xi(t_2)}_\xi [\slr_\uI (t_1), \slr_\uI (\tau_2)]\slr_\uI (t_2) \\
        &\phantom{={}}
        +\frac{\varepsilon^4}{2} \left( \int_{0}^{t} \di{\tau_1} \int_0^{\tau_1} \di{t_1} \int_0^{t_1} \di{t_2} + \int_{0}^{t}
        \di{t_2} \int_0^{t_2} \di{\tau_1} \int_0^{\tau_1} \di{t_1} + \int_{0}^{t} \di{\tau_1} \int_0^{\tau_1} \di{t_2}
        \int_0^{t_2} \di{t_1} \right)\\
        &\phantom{= + \varepsilon^4/2} 
        \times \avg{\xi(t)\xi(t_1)}_\xi \avg{\xi(\tau_1) \xi(t_2)}_\xi [\slr_\uI (\tau_1), \slr_\uI (t_1)] \slr_\uI (t_2) \\
        &\phantom{={}} 
        + \frac{\varepsilon^4}{6}\int_0^t \di{t_1} \int_0^{t_1} \di{\tau_2} \int_0^{\tau_2} \di{\tau_3} \avg{\xi(t)\xi(t_1)}_\xi \avg{\xi(\tau_2)\xi(\tau_3)}_\xi \\
        &\phantom{= + \varepsilon^4/6}
        \times \left(\left[\slr_{\uI}(t_1),\left[\slr_{\uI}(\tau_2),\slr_{\uI}(\tau_3)\right]\right] + \left[\left[\slr_{\uI}(t_1),\slr_{\uI}(\tau_2)\right],\slr_{\uI}(\tau_3)\right]\right) \\
        &\phantom{={}} 
        + \frac{\varepsilon^4}{6}\int_0^{t} \di{\tau_1} \int_0^{\tau_1} \di{t_1} \int_0^{t_1} \di{\tau_3} \avg{\xi(t)\xi(t_1)}_\xi \avg{\xi(\tau_1)\xi(\tau_3)}_\xi \\
        &\phantom{={}  + \varepsilon^4/6}
        \times \left(\left[\slr_{\uI}(\tau_1),\left[\slr_{\uI}(t_1),\slr_{\uI}(\tau_3)\right]\right] + \left[\left[\slr_{\uI}(\tau_1),\slr_{\uI}(t_1)\right],\slr_{\uI}(\tau_3)\right]\right) \\
        &\phantom{={}} 
        + \frac{\varepsilon^4}{6} \int_0^{t} \di{\tau_1} \int_0^{\tau_1} \di{\tau_2} \int_0^{\tau_2} \di{t_1} \avg{\xi(t)\xi(t_1)}_\xi \avg{\xi(\tau_1)\xi(\tau_2)}_\xi \\
        &\phantom{= + \varepsilon^4/6} 
        \times \left(\left[\slr_{\uI}(\tau_1),\left[\slr_{\uI}(\tau_2),\slr_{\uI}(t_1)\right]\right] + \left[\left[\slr_{\uI}(\tau_1),\slr_{\uI}(\tau_2)\right],\slr_{\uI}(t_1)\right]\right) \\
        &\phantom{={}}
        + \frac{\varepsilon^4}{2} \left(\int_0^{t} \di{t_1} \int_0^{t_1} \di{\tau_2} \int_0^{\tau_2} \di{t_2} 
        + \int_0^{t} \di{t_1} \int_0^{t_1} \di{t_2} \int_0^{t_2} \di{\tau_2} + \int_0^{t} \di{\tau_2} \int_0^{\tau_2} \di{t_1} \int_0^{t_1} \di{t_2} \right. \\
        &\phantom{= + \varepsilon^4/2}
        \left. + \int_0^{t} \di{\tau_2} \int_0^{\tau_2} \di{t_2} \int_0^{t_2} \di{t_1} + \int_0^{t} \di{t_2} \int_0^{t_2}
        \di{\tau_2} \int_0^{\tau_2} \di{t_1} + \int_0^{t}  \di{t_2} \int_0^{t_2} \di{t_1} \int_0^{t_1} \di{\tau_2} \right) \\
        &\phantom{= + \varepsilon^4/2}
        \times \avg{\xi(t)\xi(t_1)}_\xi \avg{\xi(\tau_2)\xi(t_2)}_\xi \left[\slr_{\uI}(\tau_2),\slr_{\uI}(t_1)\right] \slr_{\uI}(t_2) \\
        &\phantom{={}} 
        + \frac{\varepsilon^4}{4}\left(\int_0^{t} \di{t_1} \int_0^{t_1} \di{\tau_2} \int_0^{\tau_2} \di{\tau_3} + \int_0^{t}
        \di{\tau_3} \int_0^{\tau_3} \di{t_1} \int_0^{t_1} \di{\tau_2} + \int_0^{t} \di{t_1} \int_0^{t_1} \di{\tau_3}
        \int_0^{\tau_3} \di{\tau_2} \right) \\
        &\phantom{= + \varepsilon^4/4} 
        \times \avg{\xi(t)\xi(t_1)}_\xi \avg{\xi(\tau_2)\xi(\tau_3)}_\xi \left[\slr_{\uI}(\tau_3),\left[\slr_{\uI}(t_1), \slr_{\uI}(\tau_2)\right]\right] \\
        &\phantom{={}} 
        + \frac{\varepsilon^4}{4}\left(\int_0^{t} \di{\tau_1} \int_0^{\tau_1} \di{t_1} \int_0^{t_1} \di{\tau_3} + \int_0^{t}
        \di{\tau_3} \int_0^{\tau_3} \di{\tau_1} \int_0^{\tau_1} \di{t_1} + \int_0^{t} \di{\tau_1} \int_0^{\tau_1} \di{\tau_3}
        \int_0^{\tau_3} \di{t_1} \right) \\
        &\phantom{= + \varepsilon^4/4}
        \times \avg{\xi(t)\xi(t_1)}_\xi \avg{\xi(\tau_1)\xi(\tau_3)}_\xi \left[\slr_{\uI}(\tau_3),\left[\slr_{\uI}(\tau_1), \slr_{\uI}(t_1)\right]\right] \\
        &\phantom{={}}
        + \frac{\varepsilon^4}{4} \left( \int_0^{t} \di{\tau_1} \int_0^{\tau_1} \di{\tau_2} \int_0^{\tau_2} \di{t_1} + \int_0^{t}
        \di{\tau_2} \int_0^{\tau_2} \di{t_1} \int_0^{t_1} \di{\tau_1} + \int_0^{t} \di{\tau_2} \int_0^{\tau_2} \di{\tau_1}
        \int_0^{\tau_1} \di{t_1}\right)  \\
        &\phantom{= + \varepsilon^4/4}
        \times \avg{\xi(t)\xi(t_1)}_\xi \avg{\xi(\tau_1)\xi(\tau_2)}_\xi \left[\left[\slr_{\uI}(\tau_1),\slr_{\uI}(\tau_2)\right],\slr_{\uI}(t_1)\right] \\
        &\phantom{={}} 
        + \frac{\varepsilon^4}{6} \left( \int_0^{t} \di{\tau_1} \int_0^{\tau_1} \di{\tau_2} \int_0^{\tau_2} \di{t_1} + \int_0^{t}
        \di{\tau_1} \int_0^{\tau_1} \di{t_1} \int_0^{t_1} \di{\tau_2} + \int_0^{t} \di{t_1} \int_0^{t_1} \di{\tau_1}
        \int_0^{\tau_1} \di{\tau_2} \right.  \\
        &\phantom{= + \varepsilon^4/6} 
        \left. + \int_0^{t} \di{t_1} \int_0^{t_1} \di{\tau_2} \int_0^{\tau_2} \di{\tau_1} + \int_0^{t} \di{\tau_2} \int_0^{\tau_2}
        \di{t_1} \int_0^{t_1} \di{\tau_1} + \int_0^{t} \di{\tau_2} \int_0^{\tau_2} \di{\tau_1} \int_0^{\tau_1} \di{t_1} \right) \\ 
        &\phantom{=  + \varepsilon^4/6} 
        \left. \times \avg{\xi(t)\xi(t_1)}_\xi \avg{\xi(\tau_1)\xi(\tau_2)}_\xi \left[\slr_{\uI}(\tau_1),\left[\slr_{\uI}(\tau_2),
        \slr_{\uI}(t_1)\right]\right] \right\} \avg{\sL_\uI (t)}_\xi.
        \label{eq:finalTimeOrdMagnusEoM}
    \end{aligned}
\end{equation}
In Eq.~\eqref{eq:finalTimeOrdMagnusEoM}, the contribution proportional to $\varepsilon^2$ defines $\Xi^{(2)}(t)$, while the
contributions proportional to $\varepsilon^4$ collectively define $\Xi^{(4)}(t)$. Thus, Eq.~\eqref{eq:finalTimeOrdMagnusEoM} is
the explicit form of the time-local NMQME written compactly as Eq.~\eqref{eq:finalEOMKernel} in the main text.

\section{Channel diagnostics}
\label{sec:supp-channel-diagnostics}

This section provides the definitions and conventions used to assess the physicality and accuracy of the approximate finite-time
channels. We use the Choi representation to test complete positivity and trace preservation, and the diamond norm to quantify the
worst-case difference between an approximate channel and the reference channel obtained from stochastic simulations.

\subsection{Choi representation and physicality tests}

For a linear map $\Phi:\mathcal{B}(\mathcal{H}_\mm{in}) \to\mathcal{B}(\mathcal{H}_\mm{out})$, we use the unnormalized Choi
convention~\cite{Jamiolkowski1972,Choi1975,Wilde_2017},
\begin{equation}
    C_\Phi = \sum_{i,j=1}^{d_\mm{in}} \ketbra{i}{j}\otimes \Phi\left(\ketbra{i}{j}\right),
    \label{eq:supp-ChoiDefinition}
\end{equation}
where $\{\ket{i}\}_{i=1}^{d_\mm{in}}$ is an orthonormal basis of the input Hilbert space $\mathcal{H}_\mm{in}$. 

Equivalently, let $\ket{\beta} =\sum_{i=1}^{d_\mm{in}}\ket{i}\otimes\ket{i}$ denote the unnormalized maximally entangled vector on
two copies of the input space. The Choi matrix may then be written as
\begin{equation}
    C_\Phi = \left(\mathcal{I}_\mm{in}\otimes\Phi\right) \left(\ketbra{\beta}{\beta}\right),
    \label{eq:supp-ChoiEntangledDefinition}
\end{equation}
where $\mathcal{I}_\mm{in}$ is the identity map on $\mathcal{B}(\mathcal{H}_\mm{in})$. 

With this convention, complete positivity is equivalent to $C_\Phi \succeq 0$, while trace preservation requires
\begin{equation}
    \Tr_\mm{out}[C_\Phi]=\Id_\mm{in},
    \label{eq:supp-ChoiTPCondition}
\end{equation}
No factor of $d_\mm{in}$ appears in Eq.~\eqref{eq:supp-ChoiTPCondition} because $\ket{\beta}$ is unnormalized.

For the approximate channel at the final gate time, complete positivity is tested through the smallest Choi eigenvalue,
\begin{equation} 
    \lambda_\mm{min}(t_\ug) = \min_l \lambda_l\left[C_\Phi(t_\ug)\right]. 
    \label{eq:supp-MinimumChoiEigenvalue} 
\end{equation}
The channel is completely positive when $\lambda_\mm{min} (t_\ug) \geq 0$. Deviations from trace preservation are quantified by
\begin{equation}
    \Delta_\mm{TP}(t_\ug) = \norm{\Tr_\mm{out} \left[C_\Phi(t_\ug)\right] - \Id_\mm{in}}_1.
    \label{eq:supp-TPResidual}
\end{equation}
where $\norm{X}_1=\Tr\sqrt{X^\dagger X}$ is the trace norm. The residual $\Delta_\mm{TP}(t_\ug)$ vanishes precisely when the
channel is trace preserving. These definitions reproduce the diagnostics summarized in Eq.~\eqref{eq:MinChoiEigvalAndTPResidual}
of the main text.

\subsection{Diamond-norm channel error}

Physical admissibility does not determine the accuracy of an approximate channel. We therefore compare the approximate channel
$\Phi_\mm{A}(t_\ug)$ obtained with the NMQME with the reference channel $\Phi_\mm{E}(t_\ug)$ obtained from stochastic simulations.
Their difference defines the Hermiticity-preserving map $\Lambda(t_\ug) =\Phi_\mm{E}(t_\ug)-\Phi_\mm{A}(t_\ug)$, whose diamond
norm can be evaluated as~\cite{Aharonov1998,Wilde_2017,Watrous2018}
\begin{equation}
    \norm{\Lambda(t_\ug)}_\diamond = \max_{\hrho_{\mm{in},\mm{a}}} \norm{ \left(\Lambda (t_\ug) \otimes\mathcal{I}_\mm{a}\right) \left(\hrho_{\mm{in},\mm{a}}\right) }_1,
    \label{eq:supp-DiamondNormDefinition}
\end{equation}
Here, the maximization is over density operators $\hrho_{\mm{in},\mm{a}}$ of the input system and an ancilla, and
$\mathcal{I}_\mm{a}$ is the identity map on the ancilla. An ancilla with the same dimension as the input system is sufficient for
the optimization~\cite{Watrous2018,PuzzuoliWatrous2017}.

Using the notation of the main text, the diamond-norm channel error is
\begin{equation} 
    d_\diamond(t_\ug) = \norm{ \Phi_\mm{E}(t_\ug) - \Phi_\mm{A}(t_\ug) }_\diamond. 
    \label{eq:supp-DiamondNormError} 
\end{equation}
The trace norm $\norm{X}_1=\Tr\sqrt{X^\dagger X}$ in Eq.~\eqref{eq:supp-DiamondNormDefinition} quantifies the distinguishability
of the two resulting output states. The maximization over system--ancilla inputs includes states entangled with the ancilla and
therefore detects discrepancies in system--ancilla correlations. Consequently, $d_\diamond(t_\ug)$ is a state-independent
worst-case single-use channel error. With the unhalved convention used here, $0\leq d_\diamond(t_\ug)\leq2$.

\section{Ornstein--Uhlenbeck (OU) Process}
\label{sec:supp-ouProcess}

The Ornstein--Uhlenbeck (OU) process provides a stationary Gaussian noise model with a single correlation time. We define $\xi(t)$
by the stochastic differential equation~\cite{hanggi95,lehle18,maller09,contreras21}
\begin{equation}
\di{\xi(t)} = - \frac{\xi(t)}{\tau_\uc} \di{t} + \sigma\sqrt{\frac{2}{\tau_\uc}} \di{W_t},
    \label{eq:OUProcess}
\end{equation}
where $W_t$ is a standard Wiener process, $\tau_\uc$ is the correlation time, and $\sigma$ is the stationary standard deviation of
the noise. For the stationary initial condition $\xi(0)\sim\mathcal{N}(0,\sigma^2)$, the process is Gaussian with zero mean,
$\avg{\xi(t)}_\xi=0$, and correlation function
\begin{equation}
    C_\mm{OU}(t,s) = \avg{\xi(t) \xi(s)}_\xi = \sigma^2 \exp\left(- \frac{\abs{t-s}}{\tau_\uc} \right).
    \label{eq:supp-OUCorrelation}
\end{equation}
Thus, $\sigma^2 = \avg{\xi(t) \xi(t)}_\xi $ is the stationary variance, while $\tau_\uc$ determines the exponential loss of
temporal correlations. Fluctuations separated by times much longer than $\tau_\uc$ are therefore effectively uncorrelated.

\subsection{Estimating the Small Parameter Controlling Validity of NMQME}

We estimate the dimensionless parameter controlling the NMQME expansion from the second-order term in the equation of motion. The
second-order noise kernel contains the OU correlation function together with oscillatory interaction-picture superoperators
generated by the control. A representative matrix element of this kernel can be written in the form
\begin{equation}
    I (t) = \sigma^2 \int_0^t \di{t_1} \exp\left(\frac{i\theta t_1}{t_\ug}\right) \exp\left(\frac{-t_1}{\tau_\uc}\right) h(t_1),
    \label{eq:genericMatrixElement}
\end{equation}
where $\theta/t_\ug$ is the characteristic control frequency and $h(t)$ contains the remaining bounded system-dependent factors.
We assume that $h(t)$ varies slowly over the OU correlation time $\tau_\uc$. Its characteristic magnitude can then be factored
out, and the temporal size of the second-order noise kernel is determined by
\begin{equation}
    \begin{aligned}
        I_\mm{OU}(t) &= \int_0^t\di{t_1} \exp\left(\frac{i\theta t_1}{t_\ug}\right) \exp\left(\frac{-t_1}{\tau_\uc}\right) \\
        &= \frac{ 1- \exp\left[-\left(\frac{1}{\tau_\uc}-i \frac{\theta}{t_\ug}\right) t\right] }{\frac{1}{\tau_\uc}-i\frac{\theta}{t_\ug} }. 
    \end{aligned}
    \label{eq:supp-OUCorrelationIntegral} 
\end{equation}
The squared magnitude of this integral is
\begin{equation} 
    \abs{I_\mm{OU}(t)}^2 = \frac{ 1 + \exp\left(-2 \frac{t}{\tau_\uc}\right) - 2\exp\left(-\frac{t}{\tau_\uc}\right) \cos\left(\theta \frac{t}{t_\ug}\right) }{ \frac{1}{\tau_\uc^2} + \frac{\theta^2}{t_\ug^2}}. 
    \label{eq:supp-OUCorrelationIntegralSquared} 
\end{equation} 
For evolution times long compared with the correlation time, the terms proportional to $e^{-t/\tau_\uc}$ are exponentially
suppressed. Neglecting these endpoint contributions and taking the square root gives the characteristic temporal size of the
second-order OU noise kernel,
\begin{equation} 
    \abs{I_\mm{OU}(t)} \approx \frac{ \tau_\uc }{ \sqrt{ 1 + \frac{\theta^2}{t_\ug^2}\tau_\uc^2 } }.
    \label{eq:supp-OUIntegratedCorrelation} 
\end{equation} 
The quantity $\abs{I_\mm{OU}}$ estimates the temporal size of the inner noise-kernel integral. The correlation function
contributes the amplitude $\sigma^2$, while accumulation of the second-order generator over the gate contributes a factor of order
$t_\ug$. The dimensionless expansion parameter is therefore estimated as
\begin{equation} 
    \delta_\mm{OU} \approx \sigma^2t_\ug\abs{I_\mm{OU}} = \frac{ \sigma^2\tau_\uc t_\ug }{ \sqrt{ 1+ \frac{\theta^2}{t_\ug^2}\tau_\uc^2 } }. 
    \label{eq:smallParameterOU} 
\end{equation}
Equation~\eqref{eq:smallParameterOU} reproduces Eq.~\eqref{eq:SmallParameterOU} of the main text.

In the short-memory regime, $\theta\tau_\uc/t_\ug\ll1$, the parameter reduces to
\begin{equation} 
    \delta_\mm{OU} \sim \sigma^2\tau_\uc t_\ug. 
    \label{eq:supp-OUShortMemoryParameter} 
\end{equation}
The factor $\sigma^2\tau_\uc$ then acts as an effective noise-induced rate, while $t_\ug$ is the interval over which its effect
accumulates. The appearance of these two distinct timescales resembles the coarse graining underlying Brownian motion:~$\tau_\uc$
characterizes the persistence of individual fluctuations, whereas $t_\ug$ determines the duration over which their cumulative
effect is observed. We use $\delta_\mm{OU}$ to organize the comparison of the second- and fourth-order NMQME with the reference
stochastic simulations.

\subsection{Average Fidelity}
\label{sec:supp-avgFid}

As an additional state-averaged diagnostic, we compare the noise-averaged finite-time channel $\Phi(t_\ug)=\avg{\sL(t_\ug)}_\xi$ with the target gate 
$\hU_\uG = (\Id - i\hsigma_x)/\sqrt{2}$, which corresponds to a rotation by $\pi/2$ about the $x$ axis. For this comparison, we use the six
Pauli eigenstates $\hrho_{j,s}=(\Id+s\hat{\sigma}_j)/2$, where $j\in\{x,y,z\}$ identifies the Pauli operator and $s\in\{+,-\}$
identifies its eigenvalue. The channel maps each input state to the final density matrix
\begin{equation} 
    \hrho_{j,s}(t_\ug) = \Phi(t_\ug) \left[ \hrho_{j,s} \right]. 
    \label{eq:supp-FinalDensityMatrix} 
\end{equation}
The corresponding ideal final state is $\hU_\uG\hrho_{j,s}\hU_\uG^\dagger$. For a single-qubit channel, the average gate fidelity
can then be evaluated as~\cite{bowdrey02}
\begin{equation}
    \bar{F}_\mm{map} = \frac{1}{6} \sum_{\substack{j=x,y,z\\s=\pm}} \Tr\left[ \hU_\uG \hrho_{j,s} \hU_\uG^\dagger \hrho_{j,s} (t_\ug) \right].
    \label{eq:supp-AverageGateFidelity}
\end{equation}
For the NMQME, $\Phi(t_\ug)$ is obtained by propagating either the second- or fourth-order time-local equation. For the numerical
benchmark, we propagate the stochastic equation of motion independently for each sampled OU-noise realization and average the
resulting finite-time maps over the ensemble. We denote the channel estimated by this procedure as $\Phi_\mm{E}(t_\ug)$ and refer
to it as the stochastic reference channel. Because Eq.~\eqref{eq:supp-AverageGateFidelity} is linear in the final density matrix,
averaging the stochastic maps before evaluating the fidelity is equivalent to evaluating the fidelity for each realization and
then averaging over the sampled ensemble.

\begin{figure}[t]
  \centering
  \includegraphics[width=0.7\textwidth]{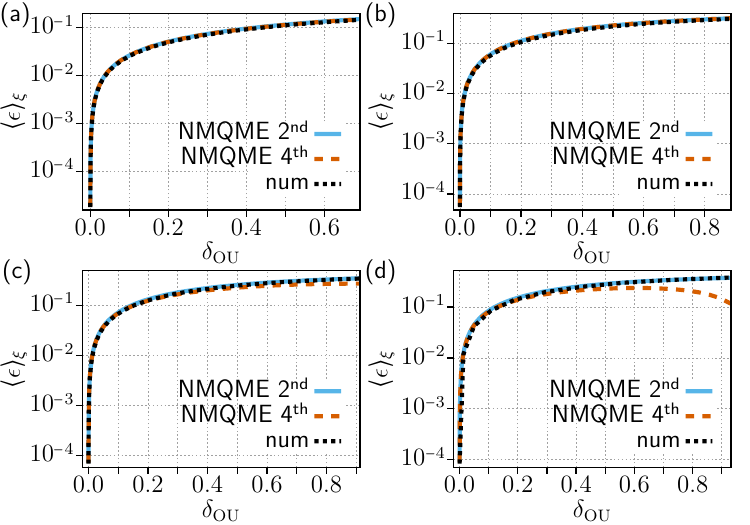}
  \caption{
      Average fidelity error $\avg{\epsilon}_{\xi}$ as a function of the small parameter $\delta_{\mathrm{OU}}$ for 
    the second-order NMQME (blue), the fourth-order NMQME (orange), and the stochastic reference channel (dashed black). The gate
    durations are
(a) $t_\ug = 1.5 \tau_\uc$, (b) $t_\ug = 3 \tau_\uc$, (c) $t_\ug = 3.5 \tau_\uc$, (d) $t_\ug = 4 \tau_\uc$.
  }
  \label{fig:supp-AvgFidOU} 
\end{figure}

We define the average fidelity error as
\begin{equation}
    \avg{\epsilon}_{\xi} = 1- \avg{\bar{F}_\mm{map}}_{\xi}.
    \label{eq:avgFid}
\end{equation}

Figure~\ref{fig:supp-AvgFidOU} compares the average fidelity error for several ratios of the gate duration to the OU
correlation time. In Figs.~\ref{fig:supp-AvgFidOU}(a) and \ref{fig:supp-AvgFidOU}(b), corresponding to $t_\ug=1.5\tau_\uc$ and
$t_\ug=3\tau_\uc$, respectively, the second- and fourth-order NMQME results closely follow the stochastic reference over the
displayed perturbative range. For the longer gate durations in Figs.~\ref{fig:supp-AvgFidOU}(c) and \ref{fig:supp-AvgFidOU}(d),
the fourth-order result develops larger deviations as $\delta_\mm{OU}$ increases, whereas the second-order result remains closer
to the stochastic reference over the displayed range.

The larger fourth-order deviation for the longer gates does not imply that the second-order approximation is generally more
accurate. Rather, it shows that adding a finite-order perturbative correction does not necessarily improve the approximation as
the parameters approach or leave the regime controlled by the expansion. The average fidelity error provides a state-averaged
diagnostic of this behavior, while the diamond-norm results in the main text compare the complete finite-time channels.

\subsection{Expectation values of Pauli operators}

The average fidelity error provides a final-time, state-averaged measure of the channel accuracy. To examine how deviations
develop during the gate, we also compare the time-resolved Bloch-vector trajectories predicted by the fourth-order NMQME with
those obtained from stochastic numerical integration. These trajectories provide a component-resolved diagnostic of the dynamics
and can reveal discrepancies that are obscured by a single final-time scalar.

For a channel $\Phi(t)$, the corresponding time-dependent density matrix is $\hrho(t) = \Phi(t) \left[ \hrho(0) \right]$. The
components of the Bloch vector $\rv$ are given by the expectation values of the Pauli operators,
\begin{equation} 
    r_j(t) = \avg{\hsigma_j (t)} = \Tr\left[ \hrho(t) \hsigma_j \right], 
    \label{eq:ExpValPauli} 
\end{equation}
with $j\in\{x,y,z\}$. Together, the three components completely determine the qubit state through $\hrho(t)= [\Id+\sum_j
r_j(t)\hsigma_j]/2$. Their time dependence therefore tests whether the NMQME reproduces the full trajectory of the noise-averaged
state, rather than only its value at the final gate time.

For the NMQME, $\Phi(t)$ is obtained by propagating the fourth-order time-local equation from the initial state, chosen to be the
excited qubit state $\hrho(0) = \ketbra{1}{1}$. For the numerical benchmark, we propagate the stochastic equation of motion
independently for each sampled OU-noise realization and average the resulting Pauli expectation values over the ensemble. The
corresponding numerical trajectories provide the stochastic reference.

We fix the dimensionless noise amplitude at $\sigma t_\ug=1$ and vary the ratio $t_\ug/\tau_\uc$. Time is displayed in units of
the OU correlation time, so each trajectory ends at $t/\tau_\uc=t_\ug/\tau_\uc$. This isolates the effect of changing the gate
duration relative to the OU memory time.

\begin{figure}[t] 
    \centering
    \includegraphics[width=0.7\textwidth]{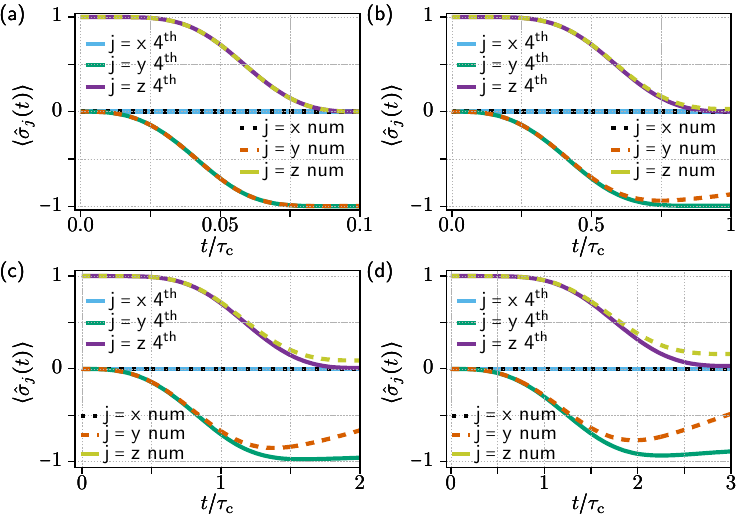}
    \caption{
        Time-resolved Pauli expectation values for OU noise, starting from the excited state $\hrho(0)=\ketbra{1}{1}$. Time is
        shown in units of the OU correlation time, $t/\tau_\uc$. The dimensionless noise amplitude is fixed at $\sigma t_\ug=1$,
        while the gate duration is varied: (a) $t_\ug/\tau_\uc=0.1$, (b) $t_\ug/\tau_\uc=1$, (c) $t_\ug/\tau_\uc=2$, and (d)
        $t_\ug/\tau_\uc=3$. The fourth-order NMQME predictions for $\avg{\hat{\sigma}_x(t)}_\xi$, $\avg{\hat{\sigma}_y(t)}_\xi$,
        and $\avg{\hat{\sigma}_z(t)}_\xi$ are shown by the blue, green, and purple solid lines, respectively. The corresponding
        stochastic-reference results are shown by the black, orange, and yellow dashed lines.
    }
    \label{fig:ExpValPauliOU}
\end{figure}

Figure~\ref{fig:ExpValPauliOU} compares the fourth-order NMQME and stochastic-reference Bloch-vector trajectories as the gate
duration is increased relative to the OU correlation time. In Fig.~\ref{fig:ExpValPauliOU}(a), the gate is short compared with the
correlation time and the expansion parameter satisfies $\delta_\mm{OU}\ll 1$. The NMQME and stochastic-reference trajectories are
nearly indistinguishable, showing that the fourth-order NMQME reproduces both the final state and the intermediate controlled
dynamics deep within the perturbative regime.

The agreement remains close for $t_\ug=\tau_\uc$ in Fig.~\ref{fig:ExpValPauliOU}(b). As $t_\ug/\tau_\uc$ increases further,
component-dependent deviations become visible in Figs.~\ref{fig:ExpValPauliOU}(c) and \ref{fig:ExpValPauliOU}(d). The largest
discrepancy occurs in the $y$ component, demonstrating that the loss of accuracy is not uniform over the Bloch vector. This
component-resolved behavior would not be apparent from the average fidelity error alone.

Figure~\ref{fig:ExpValPauliOU}(d) displays the largest overall deviations and corresponds to $\delta_\mm{OU}>1$, outside the
perturbative regime identified in the main text. The time-resolved Bloch-vector trajectories therefore support the conclusion
obtained from the channel diagnostics:~the fourth-order NMQME accurately reproduces the controlled dynamics within the
perturbative regime, while systematic and component-dependent deviations emerge as the expansion parameter approaches and exceeds
unity.

\section{1/f Noise}
\label{sec:supp-oneF}

We model low-frequency noise using a $1/f$ spectrum restricted to a finite angular-frequency interval~\cite{dutta1981},
\begin{equation}
    S(\omega) \propto \frac{1}{\omega}, \quad \omega_l \leq \omega \leq \omega_h
    \label{eq:supp-oneF-spectrum}
\end{equation}
where $\omega_l$ and $\omega_h$ are the lower and upper frequency cutoffs, respectively. The finite bandwidth removes the infrared
and ultraviolet divergences of an unrestricted $1/f$ spectrum.

Using the Wiener--Khinchin relation~\cite{papoulis2002}, the corresponding stationary correlation function is
\begin{equation} 
    \begin{aligned} 
        C_{1/f}(t,s) &= \avg{\xi(t)\xi(s)}_\xi \\ 
        &= 2\sigma^2 \left[ \Ci\left(\omega_h\abs{t-s}\right) - \Ci\left(\omega_l\abs{t-s}\right) \right], 
    \end{aligned} 
    \label{eq:corrFnOneF} 
\end{equation}
where $\sigma$ sets the amplitude of the band-limited $1/f$ spectrum and is not itself the stationary standard deviation of the
noise. The stationary variance depends on both $\sigma$ and the bandwidth, as shown below. $\Ci(t)$ denotes the cosine integral
function defined by
\begin{equation}
    \Ci(t)=-\int_t^{\infty} \di{t_1} \frac{\cos(t_1)}{t_1}.
   \label{eq:supp-CiDefinition}
\end{equation}
Although the individual cosine-integral terms diverge logarithmically as their arguments approach zero, their difference remains
finite. Using $\Ci(t)=\gamma + \ln(t) + \BigO(t^2)$ for $t \to 0$, we find
\begin{equation} 
    \avg{\xi(t)^2}_\xi = C_{1/f}(t,t) = 2\sigma^2 \ln\left( \frac{\omega_h}{\omega_l} \right). 
    \label{eq:supp-OneFVariance} 
\end{equation} 
Thus, $\sigma$ controls the overall spectral amplitude, whereas the stationary noise variance also depends logarithmically on the
bandwidth $\omega_h/\omega_l$.

In this work, we use $\omega_l/\omega_h=1/100$ for the numerical simulations. The resulting correlation function contains a broad
range of timescales rather than the single exponential correlation time characterizing OU noise.

\subsection{Estimating the Small Parameter Controlling Validity of NMQME}

Unlike OU noise, band-limited $1/f$ noise does not have a single characteristic correlation time. Instead, its temporal structure
is determined by the frequency band $\omega_l\leq\omega\leq\omega_h$. We estimate the dimensionless parameter controlling the
NMQME expansion from the second-order term in the equation of motion.

A representative matrix element of the second-order noise kernel has the form
\begin{equation} 
    I_h(t) = \int_0^t\di{t_1} \exp\left[i \theta \frac{t_1}{t_\ug}\right] C_{1/f}(t_1) h(t_1), 
    \label{eq:GenMatElFinalEoMOneF} 
\end{equation}
where $\theta/t_\ug$ is the characteristic control frequency, $C_{1/f}(t_1)$ denotes the stationary correlation function evaluated
at a time difference $t_1$, and $h(t_1)$ contains the remaining bounded system-dependent factors. We assume that $h(t_1)$ varies
slowly over the timescales relevant to the noise kernel. Its characteristic magnitude can then be factored out.

After separating the spectral-amplitude factor $\sigma^2$, the temporal size of the second-order noise kernel is determined by
\begin{equation} 
    I_{1/f}(t) = 2 \int_0^t\di{t_1} \exp\left[i \theta \frac{t_1}{t_\ug}\right] \left[ \Ci(\omega_h t_1) - \Ci(\omega_l t_1) \right]. 
    \label{eq:supp-OneFKernelIntegral} 
\end{equation}
Using the identity 
\begin{equation} 
    \Ci(\omega_h t_1) - \Ci(\omega_l t_1) = \int_{\omega_l}^{\omega_h} \frac{\di{\omega}}{\omega} \cos(\omega t_1), 
    \label{eq:supp-CiIdentity} 
\end{equation}
and evaluating the time integral at the final gate time gives
\begin{equation} 
    I_{1/f}(t_\ug) = \int_{\omega_l}^{\omega_h} \frac{\di{\omega}}{\omega} 
    \left[
    \frac{\exp\left[i \left(\frac{\theta}{t_\ug} + \omega\right) t_\ug\right] - 1}{i \left(\frac{\theta}{t_\ug} + \omega\right)} 
    + 
    \frac{\exp\left[i \left(\frac{\theta}{t_\ug} - \omega\right) t_\ug\right] - 1}{i\left(\frac{\theta}{t_\ug} - \omega\right)} 
    \right].
    \label{eq:integralOneF} 
\end{equation}
The first term in Eq.~\eqref{eq:integralOneF} is nonresonant because both $\theta/t_\ug$ and $\omega$ are positive. The real part
of the second term contains the finite-time resonant sinc kernel
\begin{equation} 
    K_{t_\ug}(\omega) = \frac{\sin\left[\left(\frac{\theta}{t_\ug}-\omega\right) t_\ug \right]}{ \frac{\theta}{t_\ug} - \omega} =
    \sinc\left[ \left(\frac{\theta}{t_\ug} - \omega\right) t_\ug \right] t_\ug, 
    \label{eq:supp-OneFResonantKernel} 
\end{equation} 
where $\sinc(x)= \sin(x)/x$. The kernel is centered at the characteristic control frequency $\omega=\theta/t_\ug$ and has a
frequency width of order $1/t_\ug$.

As $t_\ug$ increases, $K_{t_\ug}(\omega)$ becomes increasingly localized about $\omega=\theta/t_\ug$. In the distributional long-time limit,
\begin{equation} 
    K_{t_\ug}(\omega) \to \pi \delta\left( \frac{\theta}{t_\ug}-\omega \right). 
    \label{eq:supp-OneFResonantApproximation} 
\end{equation}
Provided that the characteristic control frequency lies within the noise band,
\begin{equation} 
    \omega_l < \frac{\theta}{t_\ug} < \omega_h, 
    \label{eq:supp-OneFResonanceCondition} 
\end{equation} 
the frequency integral gives the characteristic temporal size of the second-order noise kernel,
\begin{equation} 
    \abs{I_{1/f}} \approx \frac{\pi t_\ug}{\theta}. 
    \label{eq:supp-OneFIntegratedKernel} 
\end{equation} 
The correlation function contributes the spectral-amplitude factor $\sigma^2$, while accumulation of the second-order generator
over the gate contributes a further factor of order $t_\ug$. The dimensionless expansion parameter is therefore estimated as
\begin{equation}
    \delta_{1/f} \approx \sigma^2 t_\ug \abs{I_{1/f}} =  \frac{\pi}{\theta} (\sigma t_\ug)^2.
    \label{eq:supp-smallParameterOneF}
\end{equation}
Thus, $\abs{I_{1/f}}$ estimates the temporal size of the second-order noise kernel, while $\sigma^2t_\ug$ accounts for the
spectral amplitude and the accumulation of the generator over the gate. Equation~\eqref{eq:supp-smallParameterOneF} reproduces
Eq.~\eqref{eq:SmallParameterOneF} of the main text.

\subsection{Average Fidelity}
\label{sec:supp-avgFidOneF}

We evaluate the average fidelity error $\avg{\epsilon}_\xi$, defined in Eq.~\eqref{eq:avgFid}, for the band-limited $1/f$
noise model. For the NMQME, the finite-time channel $\Phi(t_\ug)$ is obtained by propagating either the second- or fourth-order
time-local equation. The stochastic reference channel is obtained by numerically propagating the stochastic equation of motion for
each sampled $1/f$-noise realization and averaging the resulting finite-time maps over the ensemble.

\begin{figure}[t]
  \centering
  \includegraphics[width=0.5\textwidth]{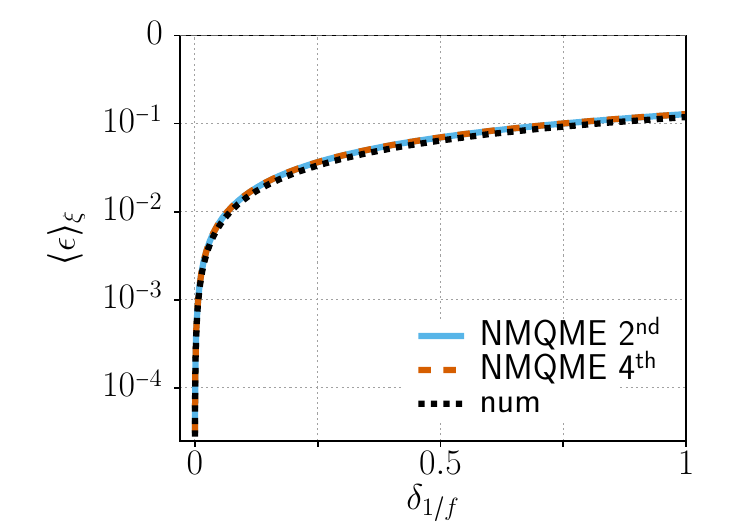}
  \caption{
      Average fidelity error $\avg{\epsilon}_\xi$ versus the expansion parameter $\delta_{1/f}$ for the second-order NMQME (blue),
      the fourth-order NMQME (orange), and the stochastic reference channel (black dashed).
  } 
  \label{fig:supp-AvgFidOneF} 
\end{figure}

Figure~\ref{fig:supp-AvgFidOneF} compares the average fidelity error as a function of $\delta_{1/f}$. The second- and
fourth-order NMQME results remain close to each other and to the stochastic reference throughout the displayed perturbative range.
In particular, the fourth-order correction produces only a small change relative to the second-order result.

This behavior is consistent with the diamond-norm comparison in the main text, where the second- and fourth-order channel errors
are also nearly indistinguishable for band-limited $1/f$ noise. The agreement between the two perturbative orders indicates that
the fourth-order contribution is comparatively small for the parameters considered here. This contrasts with the OU results, for
which the fourth-order correction produces a more visible change in the channel error. The average fidelity error therefore
supports the conclusion that the second-order NMQME already captures the dominant contribution of the band-limited $1/f$ noise
over the investigated perturbative regime.

\subsection{Expectation values of Pauli operators}
\label{sec:supp-ExpValOneF}

We complement the final-time average fidelity error with a time-resolved comparison of the Bloch-vector dynamics. Starting from
the excited qubit state $\hrho(0)=\ketbra{1}{1}$, we evaluate the Pauli expectation values defined in Eq.~\eqref{eq:ExpValPauli}
using the fourth-order NMQME and stochastic numerical integration. This component-resolved comparison tests whether the NMQME
reproduces the trajectory of the noise-averaged state throughout the gate, rather than only its final value.

For the fourth-order NMQME, the density matrix is obtained by propagating the time-local equation from the common initial state.
For the numerical benchmark, we propagate the stochastic equation of motion independently for each sampled band-limited
$1/f$-noise realization and average the resulting Pauli expectation values over the ensemble. We refer to these numerical
trajectories as the stochastic reference.

We keep the gate duration and frequency cutoffs fixed while varying the dimensionless spectral-amplitude parameter $\sigma t_\ug$.
The horizontal coordinate is the dimensionless time $\sigma t$, so each trajectory ends at $\sigma t=\sigma t_\ug$. Since
$\delta_{1/f}\propto(\sigma t_\ug)^2$, increasing the endpoint also moves the dynamics from the perturbative regime toward
stronger noise.

\begin{figure}[t] 
    \centering
    \includegraphics[width=0.7\textwidth]{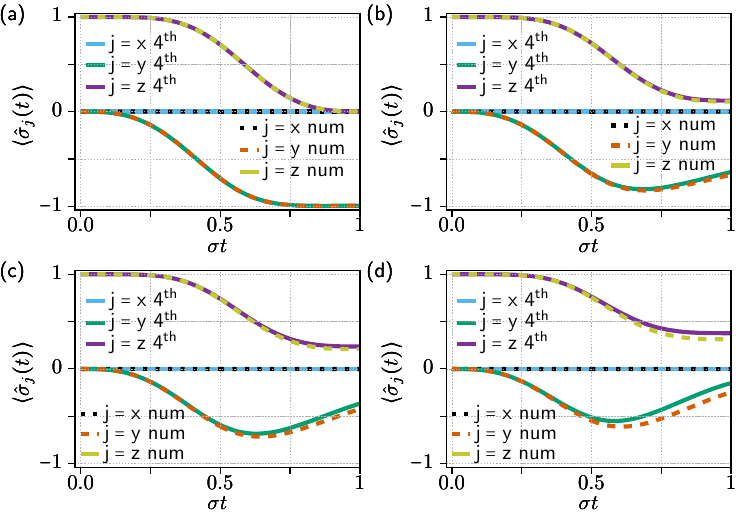}
    \caption{
        Time-resolved Pauli expectation values for band-limited $1/f$ noise, starting from the excited state
        $\hrho(0)=\ketbra{1}{1}$. Time is shown through the dimensionless coordinate $\sigma t$, so each trajectory ends at
        $\sigma t=\sigma t_\ug$.  The fourth-order NMQME predictions for $\avg{\hat{\sigma}_x(t)}_\xi$,
        $\avg{\hat{\sigma}_y(t)}_\xi$, and $\avg{\hat{\sigma}_z(t)}_\xi$ are shown by the blue, green, and purple solid lines,
        respectively. The corresponding stochastic reference results are shown by the black, orange, and yellow dashed lines. The
        dimensionless spectral amplitudes are (a) $\sigma t_\ug=0.1$, (b) $\sigma t_\ug=1$, (c) $\sigma t_\ug=1.5$, and (d)
        $\sigma t_\ug=2$.
    }
    \label{fig:ExpValPauliOneF}
\end{figure}

Figure~\ref{fig:ExpValPauliOneF} compares the fourth-order NMQME and stochastic-reference Bloch-vector trajectories for increasing
values of $\sigma t_\ug$. In Fig.~\ref{fig:ExpValPauliOneF}(a), $\sigma t_\ug=0.1$ and the expansion parameter satisfies
$\delta_{1/f} \ll 1$. The two sets of trajectories are nearly indistinguishable, showing that the fourth-order NMQME reproduces the
time-dependent state evolution within the perturbative regime.

From Fig.~\ref{fig:ExpValPauliOneF}(b) to Fig.~\ref{fig:ExpValPauliOneF}(d), increasing $\sigma t_\ug$ produces a quadratic
increase in $\delta_{1/f}$. Deviations between the fourth-order NMQME and the stochastic reference consequently become more
visible. The component-resolved trajectories also show whether the loss of accuracy is distributed uniformly over the Bloch vector
or is concentrated in particular Pauli components.

For the parameters in the later panels, $\delta_{1/f}$ approaches or exceeds unity, and the dynamics are no longer controlled by
the perturbative expansion. The increasing discrepancies are therefore consistent with the expected loss of accuracy outside the
perturbative regime. Overall, the time-resolved comparison confirms that the fourth-order NMQME captures the band-limited
$1/f$-noise dynamics when $\delta_{1/f}<1$, while systematic deviations emerge as the expansion parameter becomes large.

\section{Numerical generation and validation of the noise}
\label{sec:nv}

The stochastic reference channels used in Secs.~\ref{sec:supp-ouProcess} and~\ref{sec:supp-oneF} are obtained by propagating the
qubit under many independent realizations of the classical noise and averaging the resulting finite-time maps. The reliability of
this benchmark therefore depends on reproducing the variance, temporal covariance, and power spectrum assumed in the analytical
calculations. In this section, we summarize the numerical generators and validate their sampled statistics. Detailed algorithmic
derivations are available in the cited references and in the accompanying open-source implementation.

The qubit simulations use stationary OU noise and stationary band-limited $1/f$ noise. We also show results for the $1/f^\alpha$
generator of Kasdin~\cite{kasdin1995}. The latter is not used in the qubit calculations; it illustrates the low-frequency behavior
removed by the lower cutoff of the band-limited model.

\subsection{Conventions and numerical methods}
\label{sec:nv-conv}

Both noise models used in the qubit simulations are zero-mean Gaussian processes. A Gaussian process is fully determined by its
mean and covariance. For a stationary process, we write $C(\tau)=\avg{\xi(t+\tau)\xi(t)}_\xi$. We use the one-sided power spectral
density $S_f(f)$, with $f\geq0$, defined by
\begin{equation}
    \begin{aligned}
        C(\tau) &= \int_0^\infty\di{f} S_f(f)\cos(2\pi f\tau),
        \\
        S_f(f) &= 4\int_0^\infty\di{\tau}  C(\tau)\cos(2\pi f\tau).
    \end{aligned}
    \label{eq:nv-wk}
\end{equation}
Thus, $C(0)=\int_0^\infty\di{f}\,S_f(f)$. For spectra expressed per unit angular frequency, $\omega=2\pi f$ and $S_f(f)=2\pi
S_\omega(2\pi f)$.

Each trajectory is sampled at $t_n=n\Delta t$, $n=0,\ldots,N-1$, with record duration $T=(N-1)\Delta t$. An ensemble contains $M$
statistically independent trajectories $\xi_n^{(r)}=\xi^{(r)}(t_n)$. The OU parameter $\sigma$ is the stationary standard
deviation. For band-limited $1/f$ noise, $\sigma$ is the spectral-amplitude parameter and the stationary variance is
$C_{1/f}(0)=2\sigma^2\ln(\omega_h/\omega_l)$, as derived in Eq.~\eqref{eq:supp-OneFVariance}. The cutoffs used in the qubit
simulations satisfy $\omega_l t_\ug=1$ and $\omega_h t_\ug=100$.

\subsubsection{OU generator}
\label{sec:nv-ou}

We generate stationary OU noise using the exact sampled update of Gillespie~\cite{nv-gillespie1996}. For $\phi=\exp(-\Delta
t/\tau_\uc)$,
\begin{equation}
    \xi_{n+1} = \phi\xi_n + \sigma\sqrt{1-\phi^2} \eta_n, \quad \eta_n\sim\mathcal N(0,1),
    \label{eq:nv-ou-ar1}
\end{equation}
where the variables $\eta_n$ are independent Gaussian noise samples.  The initial value is drawn from $\xi_0\sim\mathcal
N(0,\sigma^2)$. This recursion samples the continuous OU process exactly at the grid points:
$\avg{\xi_n\xi_{n+k}}_\xi=\sigma^2\phi^{\abs{k}}$. The corresponding one-sided continuous spectrum is the Lorentzian
\begin{equation}
    S_\mm{OU}(f) = \frac{4\sigma^2\tau_\uc} {1+(2\pi f\tau_\uc)^2}.
    \label{eq:nv-ou-theory}
\end{equation}
The exact update eliminates time-discretization error in the sampled noise values. The qubit propagator and the resolved frequency
band must still be converged with respect to $\Delta t$.

\subsubsection{Band-limited $1/f$ generator}
\label{sec:nv-bl}

We generate the band-limited process using the Gaussian spectral representation of Refs.~\cite{nv-shinozuka1991,nv-timmer1995}.
The one-sided spectrum in ordinary frequency is $S_f(f)=2\sigma^2/f$ for $f_l\leq f\leq f_h$ and vanishes outside the band. A
realization is synthesized as
\begin{equation}
    \xi(t) = \sum_{k\in\mathcal K} \sqrt{w_k} \left[ a_k\cos(2\pi f_k t) + b_k\sin(2\pi f_k t) \right],
    \label{eq:nv-bl-sum}
\end{equation}
where $a_k$ and $b_k$ are independent standard normal variables.  The set $\mathcal{K}$ contains the Fourier bins that overlap the
target band, and $w_k$ is the integrated spectral power in the corresponding bin. With $f_k=k\delta f$ and $\delta f=1/(L\Delta
t)$,
\begin{equation}
    w_k = \int_{\max\{(k-\frac12)\delta f,f_l\}}^{\min\{(k+\frac12)\delta f,f_h\}} \di{f} \frac{2\sigma^2}{f}.
    \label{eq:nv-bl-w}
\end{equation}
The bin-integrated weights give the exact target variance. The sampled covariance of the finite synthesis is
\begin{equation}
    C_\mm{gen}(\tau) = \sum_{k\in\mathcal K} w_k\cos(2\pi f_k\tau).
    \label{eq:nv-bl-Cgen}
\end{equation}
The synthesis is evaluated by an inverse real FFT. We take $L\geq8N$ and retain the first $N$ samples to keep the analyzed record
well inside one period of the finite Fourier representation. For the reference ensemble, $L=\numBlL$ and $\abs{\mathcal
K}=\numBlNbins$. The maximum relative difference between $C_\mm{gen}$ and the continuum cosine-integral covariance over the
analyzed lags is $\numBlCiMaxdiff$.

\subsubsection{Kasdin generator without a lower cutoff}
\label{sec:nv-kasdin}

For comparison, we use the fractional-difference construction of Kasdin~\cite{kasdin1995,nv-hosking1981}. The filter
$H(z)=(1-z^{-1})^{-\alpha/2}$ is applied to independent Gaussian noise samples of variance $Q_d=Q\Delta t^{\alpha-1}$. The filter
coefficients obey
\begin{equation}
    \begin{aligned}
        h_0 &=1,\\
        h_k &= \frac{k-1+\alpha/2}{k}\,h_{k-1}.
    \end{aligned}
    \label{eq:nv-k-h}
\end{equation}
The numerical sequence is initialized with zero prehistory and formed by the convolution
\begin{equation}
    \xi_n = \sum_{k=0}^{n}h_k w_{n-k}.
    \label{eq:nv-k-convolution}
\end{equation}
Its finite-record covariance is known exactly,
\begin{equation}
    \Sigma_{nm} = Q_d \sum_{s=0}^{\min(n,m)} h_{n-s}h_{m-s}.
    \label{eq:nv-k-cov}
\end{equation}
For $\alpha=1$, the variance grows logarithmically and no finite-variance stationary state exists. The convolution is evaluated
with zero-padded FFTs. This generator is included only to demonstrate the infrared behavior regularized by the lower cutoff of the
band-limited model.

\subsection{Statistical tests}
\label{sec:nv-method}

At each time $t_n$, the ensemble mean and variance are estimated as
\begin{equation}
    \begin{aligned}
        \widehat{\mu}(t_n) &= \frac{1}{M}\sum_{r=1}^{M}\xi_n^{(r)}, \\
        \widehat{\sigma}^2(t_n) &= \frac{1}{M-1} \sum_{r=1}^{M} \left[\xi_n^{(r)}-\widehat{\mu}(t_n)\right]^2.
    \end{aligned}
    \label{eq:nv-ensembleMoments}
\end{equation}
Because the theoretical mean is known to vanish, we do not subtract the time average of each individual trajectory when estimating
its autocovariance. The trajectory-level and ensemble-averaged estimators are
\begin{equation}
    \begin{aligned}
        \widehat C^{(r)}(\tau_k) &= \frac{1}{N-k} \sum_{n=0}^{N-1-k} \xi_n^{(r)}\xi_{n+k}^{(r)}, \\
        \widehat C(\tau_k) &= \frac{1}{M}\sum_{r=1}^{M}\widehat C^{(r)}(\tau_k).
    \end{aligned}
    \label{eq:nv-acf}
\end{equation}
For the one-sided periodogram, define $X_k^{(r)}=\sum_{n=0}^{N-1}\xi_n^{(r)}e^{-i2\pi kn/N}$ and $f_k=k/(N\Delta t)$. Then
\begin{equation}
    \begin{aligned}
        \widehat S^{(r)}(f_k) &= g_k\frac{\Delta t}{N}\abs{X_k^{(r)}}^2, \\
        \widehat S(f_k) &= \frac{1}{M}\sum_{r=1}^{M}\widehat S^{(r)}(f_k),
    \end{aligned}
    \label{eq:nv-pgram}
\end{equation}
where $g_k=2$ for interior positive-frequency bins and $g_{N/2}=1$ for the Nyquist bin when $N$ is even.

Finite records modify the expected periodogram and, for nonstationary noise, the expected time-averaged autocovariance. We
therefore compare each estimate with its exact finite-record expectation computed from the covariance matrix
$\Sigma_{nm}=\avg{\xi_n\xi_m}_\xi$:
\begin{equation}
    \begin{aligned}
        \avg{\widehat C(\tau_k)}_\xi &= \frac{1}{N-k} \sum_{n=0}^{N-1-k}\Sigma_{n,n+k}, \\
        \avg{\widehat S(f_k)}_\xi &= g_k\frac{\Delta t}{N} \sum_{n,m=0}^{N-1} \Sigma_{nm}e^{-i2\pi k(n-m)/N}.
    \end{aligned}
    \label{eq:nv-finiteRecordExpectations}
\end{equation}
Standard errors are obtained from the variation of the independent trajectory-level estimates. For each tested quantity, we form
the standardized residual
\begin{equation}
    z_i = \frac{\widehat Q_i-\avg{\widehat Q_i}_\xi} {\mm{SE}_i}.
    \label{eq:nv-standardizedResidual}
\end{equation}
When $n_\mm{test}$ values are examined simultaneously, we use the Bonferroni threshold
\begin{equation}
    z^\ast_{n_\mm{test}} = \sqrt{2} \operatorname{erfc}^{-1} \left(\frac{10^{-3}}{n_\mm{test}}\right),
    \label{eq:nv-bonf}
\end{equation}
corresponding to a family-wise false-alarm probability no greater than $10^{-3}$ under the standard-normal approximation
~\cite{nv-dunn1961}. We test the final-time mean and variance, the autocovariance over all analyzed lags, and the periodogram over
all analyzed frequencies.

\begin{table}[t]
    \centering
    \begin{tabular}{llcccc}
        \hline
        \hline
        Generator & model parameters & $\Delta t/t_0$ & $N$ & $M$ & $T/t_0$\\
        \hline
        OU & $\sigma=\numOuSigma$, $\tau_\uc=\numOuTauC$ & \numOuDt & \numOuN & \numOuM & \numOuT\\
        band-limited $1/f$ & $\sigma=\numBlSigma$, $\omega_l t_\ug=1$, $\omega_h t_\ug=100$ & \numBlDt & \numBlN & \numBlM & \numBlT\\
        Kasdin $1/f^\alpha$ & $\alpha=\numPinkAlpha$, $Q=\numPinkQ$ & \numPinkDt & \numPinkN & \numPinkM & \numPinkT\\
        \hline
        \hline
    \end{tabular}
    \caption{Parameters of the numerical reference ensembles. Times are in units of $t_0=\tau_\uc$ (OU) and $t_0=t_\ug$ ($1/f$).}
    \label{tab:nv-params}
\end{table}

\begin{table}[t]
    \centering
    \begin{tabular}{lcccc}
        \hline
        \hline
        Generator & mean & variance & autocovariance & periodogram\\
        \hline
        OU & \numOuZMean\ (\numOuZlimMean) & \numOuZVar\ (\numOuZlimVar) & \numOuZAcov\ (\numOuZlimAcov) & \numOuZPsd\ (\numOuZlimPsd)\\
        band-limited $1/f$ & \numBlZMean\ (\numBlZlimMean) & \numBlZVar\ (\numBlZlimVar) & \numBlZAcov\ (\numBlZlimAcov) & \numBlZPsd\ (\numBlZlimPsd)\\
        Kasdin $1/f$ & \numPinkZMean\ (\numPinkZlimMean) & \numPinkZVar\ (\numPinkZlimVar) & \numPinkZAcov\ (\numPinkZlimAcov) & \numPinkZPsd\ (\numPinkZlimPsd)\\
        \hline
        \hline
    \end{tabular}
    \caption{Largest absolute standardized residual for each validation test; the corresponding Bonferroni threshold is shown in parentheses.}
    \label{tab:nv-summary}
\end{table}

\subsection{Results: Ornstein--Uhlenbeck noise}
\label{sec:nv-res-ou}

\begin{figure}[t]
    \centering
    \includegraphics[width=\textwidth]{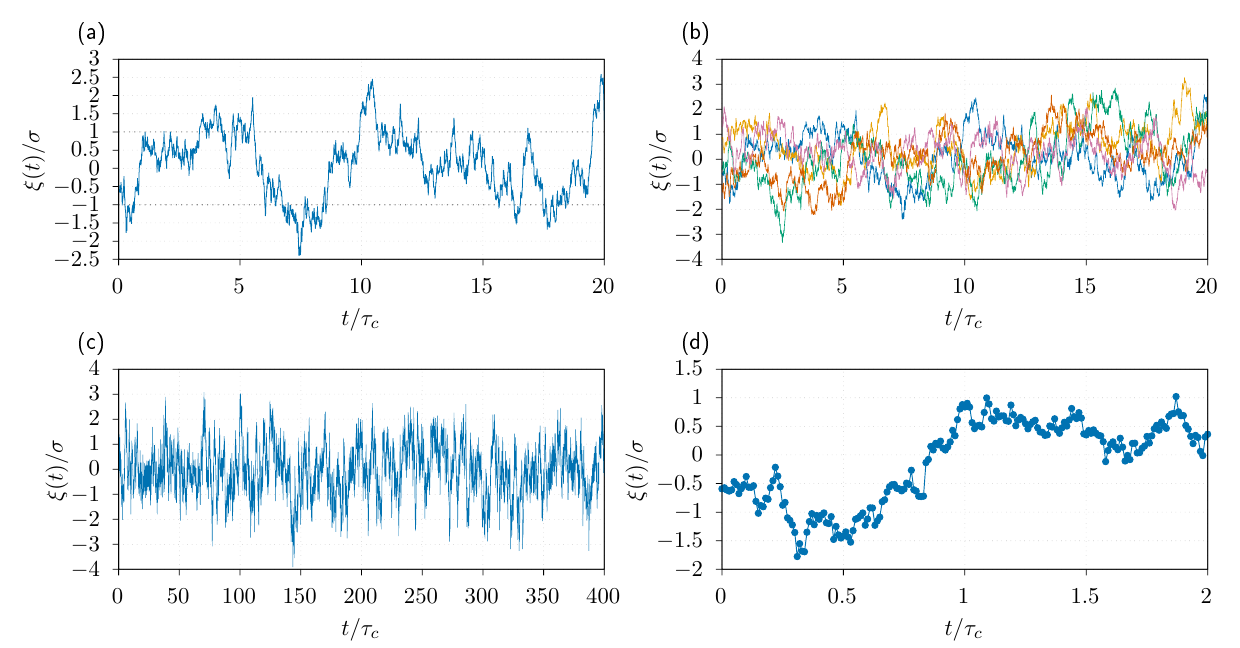}
    \caption{
        Representative OU trajectories generated with Eq.~\eqref{eq:nv-ou-ar1}. Noise and time are shown as $\xi(t)/\sigma$ and
        $t/\tau_\uc$, respectively, with $\Delta t/\tau_\uc=\numOuDt$. (a) One trajectory. (b) Five independent trajectories. (c)
        A trajectory of duration $400\tau_\uc$. (d) The first $2\tau_\uc$ of panel~(a), with the individual samples marked.
    }
  \label{fig:nv-ou-paths}
\end{figure}

\begin{figure}[t]
    \centering
    \includegraphics[width=\textwidth]{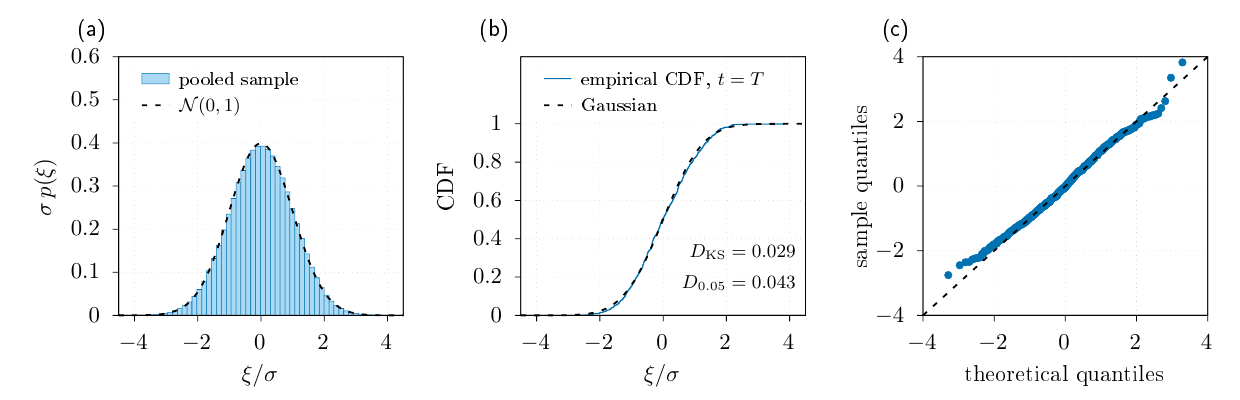}
    \caption{
        Gaussian marginal statistics of the OU noise. (a) Histogram of the normalized samples $\xi/\sigma$ and the standard normal
        target density. (b) Empirical cumulative distribution of the normalized final-time cross-section and the standard normal
        cumulative distribution. (c) Quantile--quantile plot against the standard normal distribution; the dashed line is the
        identity.
    }
    \label{fig:nv-ou-dist}
\end{figure}

Figures~\ref{fig:nv-ou-paths} and~\ref{fig:nv-ou-dist} show representative trajectories and the stationary Gaussian marginal. The
histogram of all samples is descriptive because samples within a trajectory are correlated. The quantitative Gaussianity test uses
the final-time cross-section, containing one independent value from each trajectory. Its Kolmogorov--Smirnov distance is
$D_\mm{KS}=\numOuKsD$, below the $5\%$ critical value $\numOuKsCrit$. The skewness and excess kurtosis are
$\numOuSkew\pm\numOuSkewSe$ and $\numOuKurt\pm\numOuKurtSe$, respectively, and are consistent with zero.

\begin{figure}[t]
    \centering
    \includegraphics[width=\textwidth]{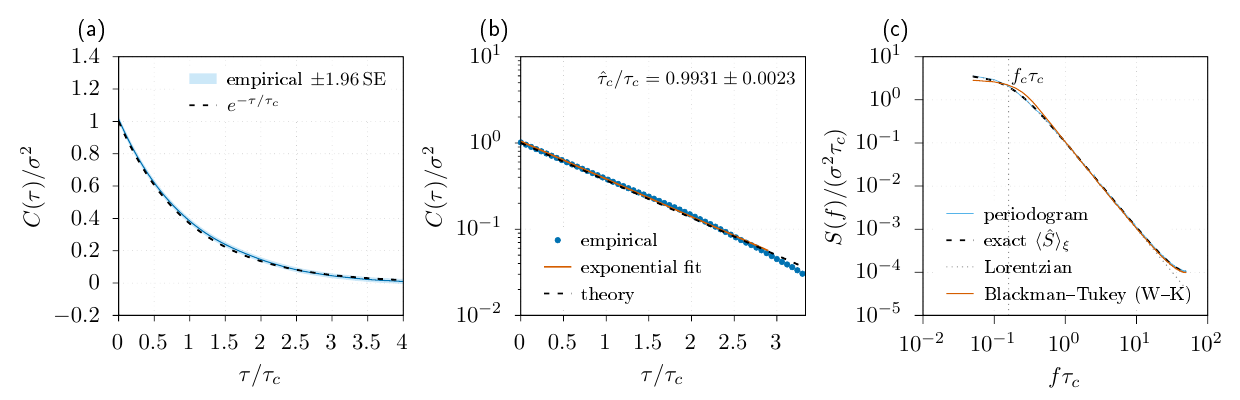}
    \caption{
        Second-order OU statistics from $M=\numOuM$ trajectories. Lag, autocovariance, frequency, and spectral density are
        displayed as $\tau/\tau_\uc$, $C(\tau)/\sigma^2$, $f\tau_\uc$, and $S(f)/(\sigma^2\tau_\uc)$, respectively. (a)
        Autocovariance with its pointwise $\pm1.96\,\mm{SE}$ band and the target covariance from
        Eq.~\eqref{eq:supp-OUCorrelation}. (b) The same data on a logarithmic vertical axis and the fitted correlation time. (c)
        Periodogram with its exact finite-record expectation (dashed), the continuous Lorentzian (gray dotted), and the
        Blackman--Tukey estimate. The vertical dotted line marks $f_\uc\tau_\uc=\numOuFCorner$.
    }
  \label{fig:nv-ou-acf}
\end{figure}

Figure~\ref{fig:nv-ou-acf}(a) compares the estimated autocovariance with Eq.~\eqref{eq:supp-OUCorrelation}. The shaded region is a
pointwise $95\%$ band; neighboring lags are correlated, so the global comparison is made using Eq.~\eqref{eq:nv-bonf}. The largest
standardized residual is $\numOuAcfMaxz$, below the threshold in Table~\ref{tab:nv-summary}. The log-linear fit in panel (b) gives
$\widehat{\tau}_\uc/\tau_\uc=\numOuTauCHat\pm\numOuTauCHatSe$, compared with the exact dimensionless value
$\widehat{\tau}_\uc/\tau_\uc=1$. The replicated analysis below shows that the single-fit regression error underestimates the full
sampling uncertainty.

In Fig.~\ref{fig:nv-ou-acf}(c), the periodogram follows its exact finite-record expectation, with maximum standardized residual
$\numOuPsdMaxz$. The continuous Lorentzian is shown separately. The flattening near the Nyquist frequency is the expected
consequence of aliasing. The Blackman--Tukey spectrum, obtained from the Parzen-windowed autocovariance, provides a consistent
time-domain estimate of the same spectrum.

\subsection{Results: band-limited $1/f$ noise}
\label{sec:nv-res-bl}

\begin{figure}[t]
    \centering
    \includegraphics[width=\textwidth]{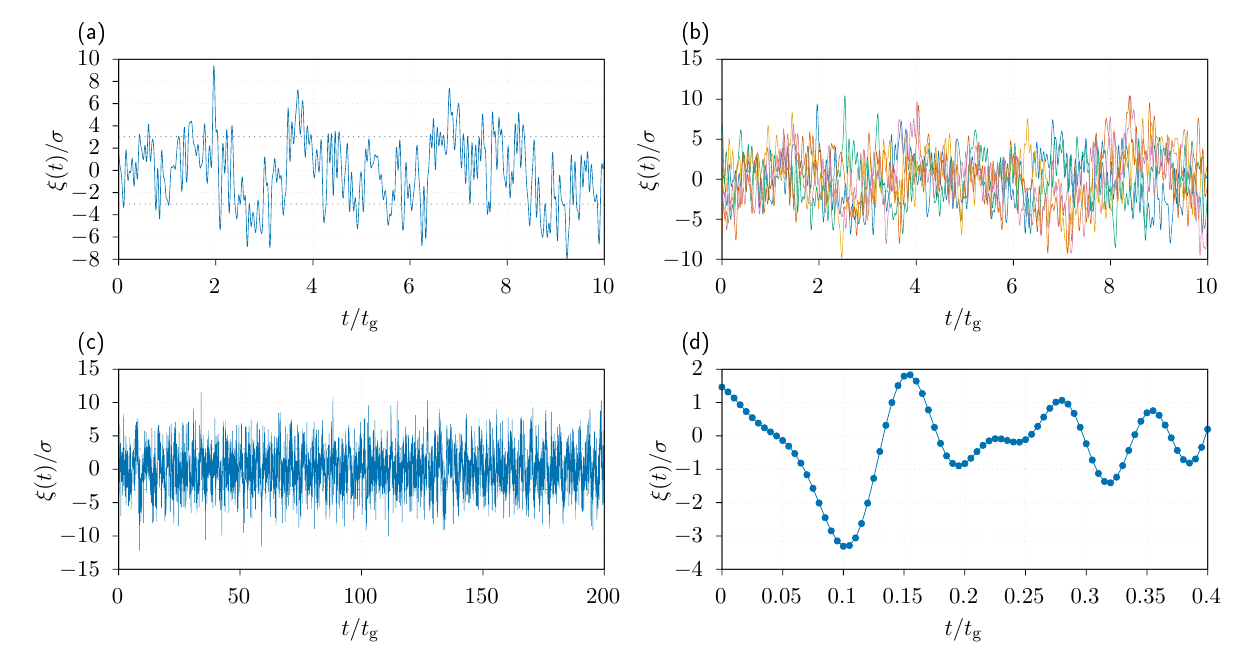}
    \caption{
        Representative band-limited $1/f$ trajectories generated by Eq.~\eqref{eq:nv-bl-sum}. Noise and time are displayed as
        $\xi(t)/\sigma$ and $t/t_\ug$, respectively. The cutoffs satisfy $\omega_l t_\ug=1$ and $\omega_h t_\ug=100$. (a) One
        trajectory, with dotted lines at $\pm\sqrt{C_{1/f}(0)}/\sigma$. (b) Five independent trajectories. (c) A trajectory of
        duration $\numBlTLong\,t_\ug$. (d) The first $\numBlTZoom\,t_\ug$ of panel~(a).
    }
    \label{fig:nv-bl-paths}
\end{figure}

\begin{figure}[t]
    \centering
    \includegraphics[width=\textwidth]{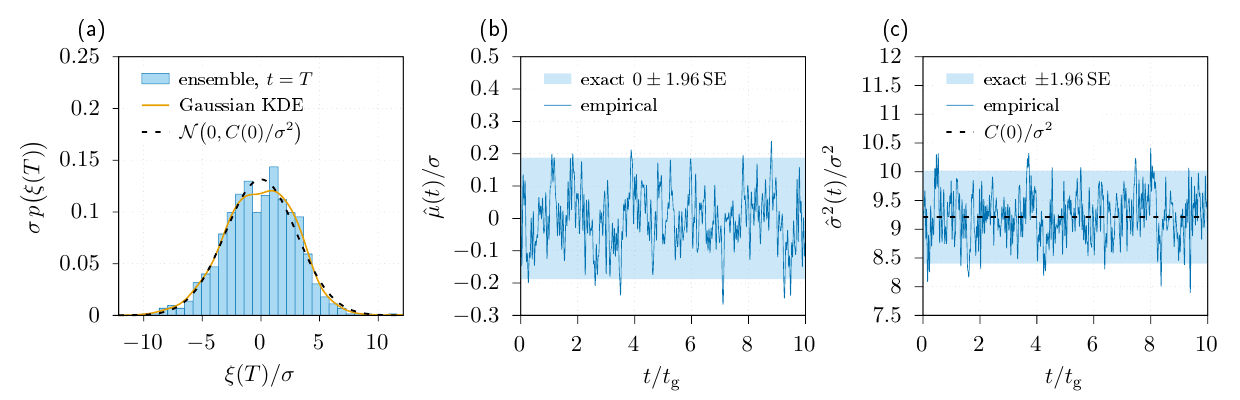}
    \caption{
        One-time statistics of the band-limited $1/f$ ensemble. Time is shown as $t/t_\ug$; noise, ensemble mean, and ensemble
        variance are normalized by $\sigma$, $\sigma$, and $\sigma^2$, respectively. (a) Final-time probability density and the
        corresponding Gaussian target distribution. (b) Ensemble mean with its pointwise $95\%$ uncertainty band. (c) Ensemble
        variance with its pointwise $95\%$ uncertainty band about $C_{1/f}(0)/\sigma^2 =2\ln(\omega_h/\omega_l)$.
    }
    \label{fig:nv-bl-dist}
\end{figure}

Figures~\ref{fig:nv-bl-paths} and~\ref{fig:nv-bl-dist} show representative trajectories and one-time statistics. The final-time
cross-section is consistent with a Gaussian distribution of variance $C_{1/f}(0)$; the empirical variance is $\numBlVarTEmp$,
compared with the target $\numBlVarExact$. The ensemble mean fluctuates around zero and the variance around a time-independent
value. The displayed bands are pointwise intervals, and neighboring times are correlated. The absence of systematic variance
growth is consistent with stationarity.

\begin{figure}[t]
    \centering
    \includegraphics[width=\textwidth]{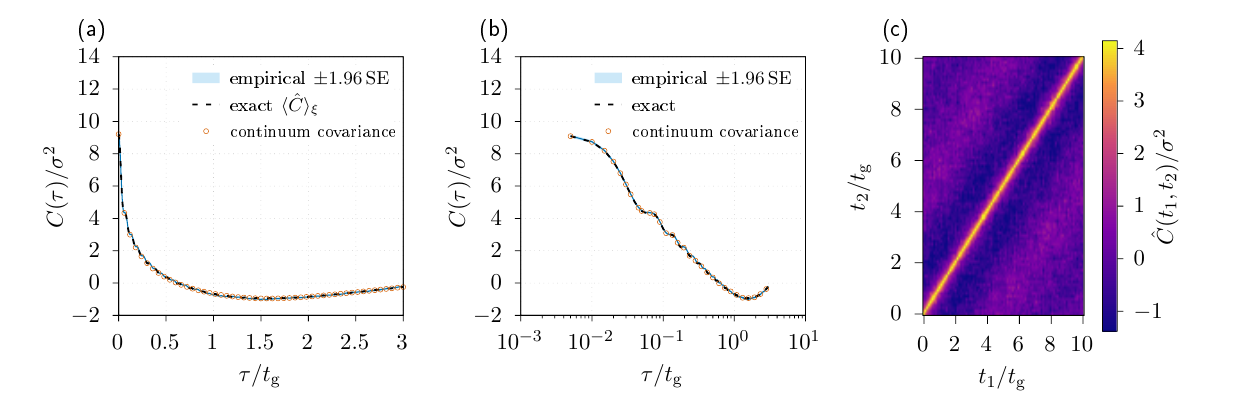}
    \caption{
        Two-time statistics of the band-limited $1/f$ ensemble. Covariance is normalized by $\sigma^2$, and lag and time are shown
        in units of $t_\ug$. (a) Autocovariance with its pointwise uncertainty band, the exact finite-grid covariance $C_\mm{gen}$
        (dashed), and the continuum covariance from Eq.~\eqref{eq:corrFnOneF} (dotted). (b) The same data on a logarithmic lag
        axis. (c) Ensemble two-time covariance on a $90\times90$ time grid.
    }
    \label{fig:nv-bl-acf}
\end{figure}

Figure~\ref{fig:nv-bl-acf} compares the measured autocovariance with $C_\mm{gen}(\tau)$ and with the continuum covariance in
Eq.~\eqref{eq:corrFnOneF}. The two references differ by at most $\numBlCiMaxdiff\,C_{1/f}(0)$ over the analyzed lags. The measured
covariance follows its finite-grid expectation with maximum standardized residual $\numBlAcfMaxz$. The two-time covariance depends
only on the time difference, as required for stationarity; its maximum elementwise standardized residual is $\numBlTwotimeMaxz$,
below $\numBlTwotimeZlim$.

\begin{figure}[t]
    \centering
    \includegraphics[width=\textwidth]{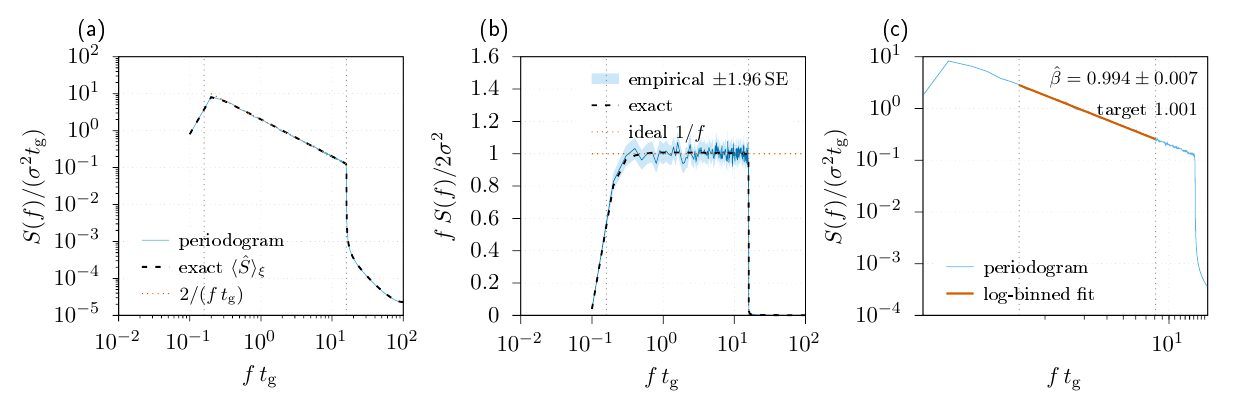}
    \caption{
        Spectrum of the band-limited $1/f$ ensemble. Frequency and spectral density are displayed as $f t_\ug$ and
        $S(f)/(\sigma^2t_\ug)$, respectively. (a) Periodogram, exact finite-record expectation (dashed), and target $1/f$ spectrum
        (dotted). (b) Compensated spectrum $f\widehat S(f)/(2\sigma^2)$ with its pointwise uncertainty band. (c) Log-binned
        power-law fit. The vertical dotted lines mark $f_l t_\ug=\numBlFl$ and $f_h t_\ug=\numBlFh$.
    }
    \label{fig:nv-bl-psd}
\end{figure}

Figure~\ref{fig:nv-bl-psd} compares the periodogram with its exact finite-record expectation. Within the prescribed band the
spectrum follows the target $1/f$ dependence. Outside the band, the ideal spectrum vanishes but the finite-record periodogram has
the leakage floor reproduced by the exact expectation. The maximum standardized residual is $\numBlPsdMaxz$. The compensated
spectrum is approximately flat at unity in the band interior. The log-binned fit gives
$\widehat\beta=\numBlBetaHat\pm\numBlBetaSe$, compared with the finite-record target $\numBlBetaTarget$.

\subsection{Results: $1/f$ noise without a lower cutoff}
\label{sec:nv-res-kasdin}

\begin{figure}[t]
    \centering
    \includegraphics[width=\textwidth]{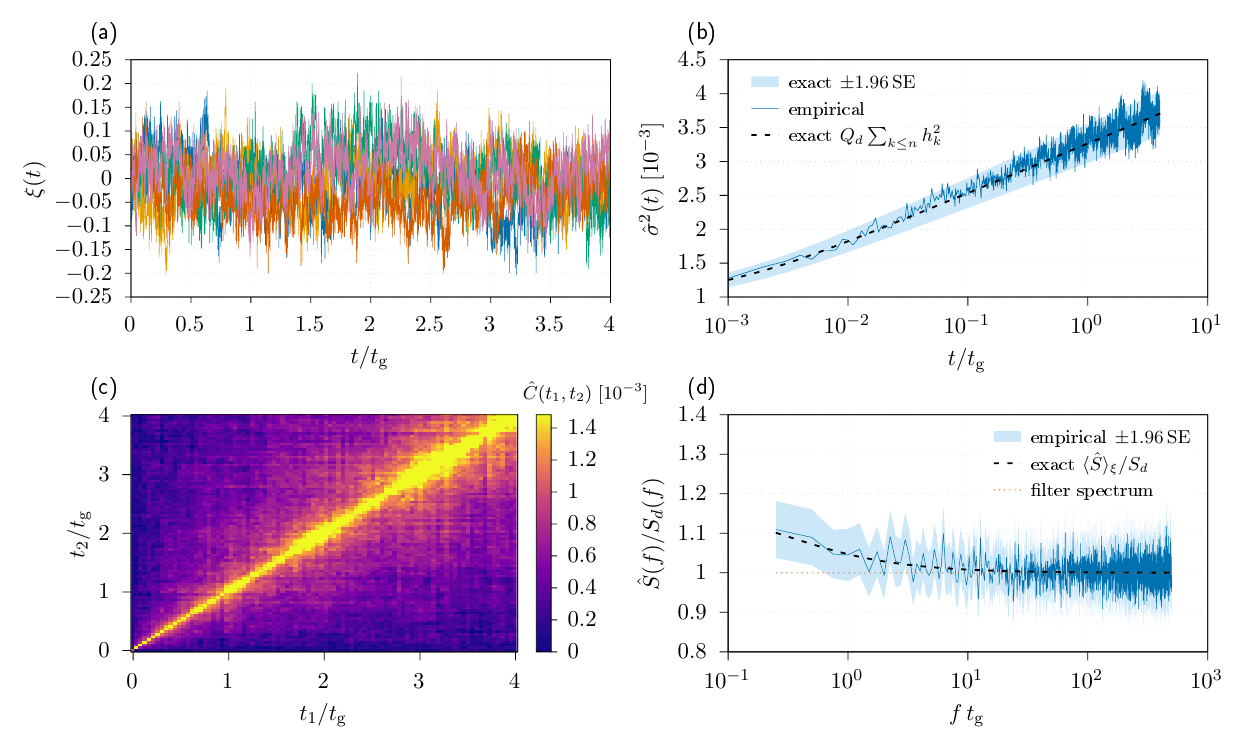}
    \caption{
        Finite-record statistics of the Kasdin $1/f^\alpha$ generator with $\alpha=\numPinkAlpha$, $Q=\numPinkQ$, and
        $M=\numPinkM$. Time and frequency are shown as $t/t_\ug$ and $f t_\ug$, respectively, to facilitate comparison with the
        band-limited $1/f$ process. (a) Five trajectories. (b) Ensemble variance and exact time-dependent expectation. (c)
        Ensemble two-time covariance. (d) Periodogram and exact finite-record expectation, both divided by the fractional-filter
        spectrum. Shaded regions are pointwise $95\%$ bands.
    }
    \label{fig:nv-kasdin}
\end{figure}

Figure~\ref{fig:nv-kasdin} illustrates the infrared behavior removed by the lower cutoff of the band-limited model. The variance
grows from $Q_d=\numPinkVarZeroExact$ at the beginning of the record to the exact finite-record value $\numPinkVarTExact$ at $T$,
compared with the sample value $\numPinkVarTEmp$. The two-time covariance is not Toeplitz and agrees with Eq.~\eqref{eq:nv-k-cov};
its maximum standardized residual is $\numPinkTwotimeMaxz$, below $\numPinkTwotimeZlim$.

The periodogram in panel (d) is divided by the fractional-filter spectrum. At higher frequencies the ratio approaches unity. At
the lowest frequencies, the exact finite-record periodogram exceeds the filter spectrum because the sequence is initialized with
zero prehistory and has a time-dependent covariance. The ratio in the lowest bin is $\numPinkLowbinExcess$, and the maximum
standardized residual is $\numPinkPsdMaxz$. Thus, the growing variance and low-frequency excess are properties of the finite
unrestricted construction, not numerical failures.

\subsection{Comparison of OU and band-limited $1/f$ noise}
\label{sec:nv-cmp}

\begin{figure}[t]
    \centering
    \includegraphics[width=\textwidth]{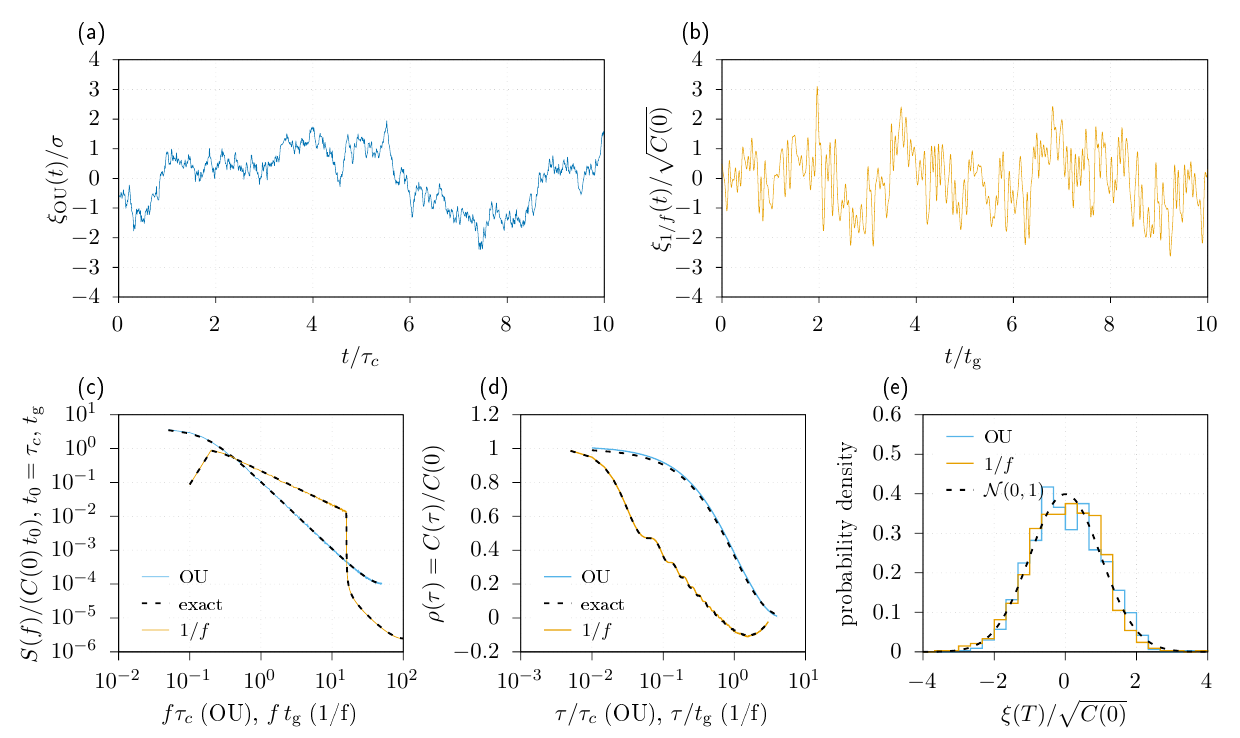}
    \caption{
        Comparison of the OU and band-limited $1/f$ noise models after normalization by their stationary standard deviations. (a)
        A normalized OU trajectory, $\xi(t)/\sigma$, as a function of $t/\tau_\uc$, showing fluctuations governed by the single
        correlation time $\tau_\uc$. (b) A normalized band-limited $1/f$ trajectory, $\xi(t)/\sqrt{C_{1/f}(0)}$, as a function of
        $t/t_\ug$, showing the coexistence of fluctuations over the range of timescales set by $\omega_l t_\ug=1$ and $\omega_h
        t_\ug=100$. (c) One-sided periodograms divided by the corresponding stationary variance, with their exact finite-record
        expectations shown as dashed curves. Frequency is displayed as $f\tau_\uc$ for OU noise and $f t_\ug$ for $1/f$ noise. (d)
        Normalized autocovariances $\widehat{C}(\tau)/C(0)$ as functions of $\tau/\tau_\uc$ for OU noise and $\tau/t_\ug$ for
        $1/f$ noise; dashed curves show the corresponding exact expectations. (e) Histograms of the normalized final-time
        cross-sections, compared with the standard normal density $\mathcal{N}(0,1)$.
    }
    \label{fig:nv-cmp}
\end{figure}

Figure~\ref{fig:nv-cmp} compares the two stationary Gaussian models after normalization by their stationary standard deviations.
Their standardized one-time distributions are identical, but their temporal correlations differ. OU noise has a single correlation
time and a Lorentzian spectrum. Band-limited $1/f$ noise combines fluctuations across the full prescribed frequency band and
retains appreciable correlations over a broader range of lags. The comparison demonstrates why matching only the variance and
marginal distribution is insufficient when validating a noise generator.

\subsection{Convergence with ensemble size}
\label{sec:nv-convergence}

To assess Monte Carlo convergence and the uncertainty of fitted quantities, we repeat the complete estimation procedure on
independent groups drawn from pools of $\numConvPool$ trajectories. For each $M\in\{10,25,50,100,200,400,1000\}$, the pool is
divided into up to $\numConvRmax$ disjoint groups. 

\begin{figure}[t]
    \centering
    \includegraphics[width=\textwidth]{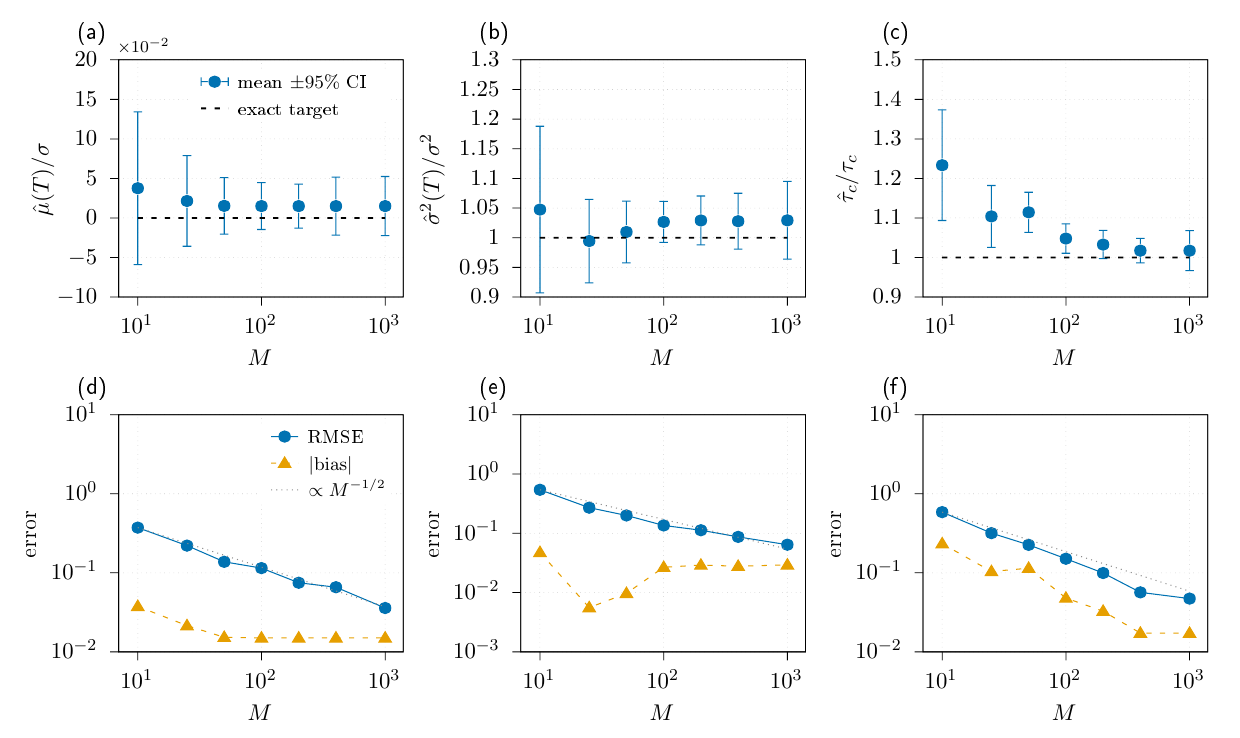}
    \caption{
        Convergence of the OU estimators with ensemble size $M$. (a) Mean final-time estimate $\overline{\widehat{\mu}(T)}/\sigma$
        and its $95\%$ confidence interval; the dashed line marks the exact value zero. (b) Mean final-time variance estimate
        $\overline{\widehat{\sigma}^2(T)}/\sigma^2$ and its $95\%$ confidence interval; the dashed line marks the exact value one.
        (c) Mean fitted correlation-time ratio $\overline{\widehat{\tau}_\uc}/\tau_\uc$ and its $95\%$ confidence interval; the
        dashed line marks the exact value one. (d)--(f) Root-mean-square errors and absolute biases corresponding to the
        estimators in panels (a)--(c), respectively. The dotted guides show the expected $M^{-1/2}$ Monte Carlo scaling.
    }
    \label{fig:nv-conv-ou}
\end{figure}

\begin{figure}[t]
    \centering
    \includegraphics[width=\textwidth]{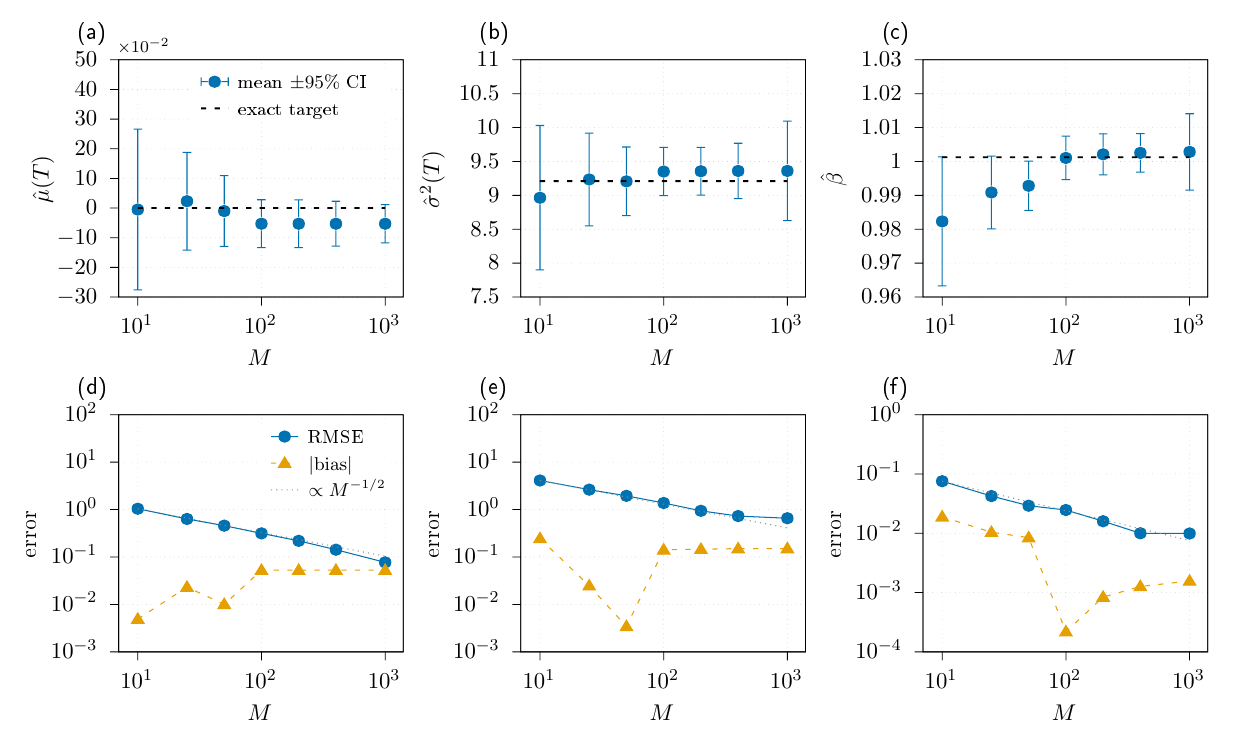}
    \caption{
        Convergence of the band-limited $1/f$ estimators with ensemble size $M$. (a) Mean final-time estimate
        $\overline{\widehat{\mu}(T)}/\sigma$ and its $95\%$ confidence interval; the dashed line marks the exact value zero. (b)
        Mean final-time variance estimate $\overline{\widehat{\sigma}^2(T)}/\sigma^2$ and its $95\%$ confidence interval; the
        dashed line marks the exact value $C_{1/f}(0)/\sigma^2 =2\ln(\omega_h/\omega_l)$. (c) Mean fitted spectral exponent
        $\overline{\widehat{\beta}}$ and its $95\%$ confidence interval; the dashed line is obtained by applying the same fit to
        the exact finite-record periodogram. (d)--(f) Root-mean-square errors and absolute biases corresponding to the estimators
        in panels (a)--(c), respectively. The dotted guides show the expected $M^{-1/2}$ Monte Carlo scaling.
    }
    \label{fig:nv-conv-bl}
\end{figure}

Figures~\ref{fig:nv-conv-ou}(a)--\ref{fig:nv-conv-ou}(c) and \ref{fig:nv-conv-bl}(a)--\ref{fig:nv-conv-bl}(c) show the means of
the replicate estimates with Student-$t$ confidence intervals. The corresponding RMSEs and absolute biases are shown in
Figs.~\ref{fig:nv-conv-ou}(d)--\ref{fig:nv-conv-ou}(f) and \ref{fig:nv-conv-bl}(d)--\ref{fig:nv-conv-bl}(f), respectively. The
dotted lines indicate the expected $M^{-1/2}$ Monte Carlo scaling.

The OU mean and variance converge at the expected rate. The fitted correlation time is biased upward for small ensembles because
it is a nonlinear function of the noisy autocovariance. Replicating the full fit gives a more realistic uncertainty than the
regression error of a single fit. For band-limited $1/f$ noise, the mean and variance show the same Monte Carlo scaling, while the
fitted spectral exponent is biased downward for small $M$. The bias decreases as the periodogram becomes less noisy. These biases
arise from the nonlinear estimators, not from the generators, which pass the parameter-free covariance and spectral tests in
Table~\ref{tab:nv-summary}.

\subsection{Computational cost}
\label{sec:nv-cost}

\begin{figure}[t]
    \centering
    \includegraphics[width=\textwidth]{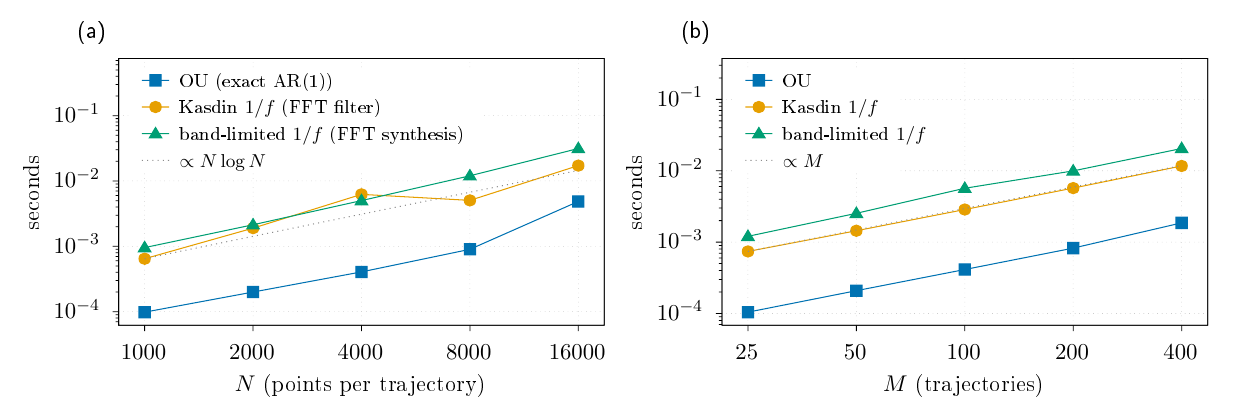}
    \caption{
        Wall-clock time to generate a complete ensemble, using the best of three runs after a warm-up run. (a) Time versus
        retained record length $N$ at $M=50$. (b) Time versus ensemble size $M$ at $N=2001$. Dotted guides show $N\log N$ and $M$
        scaling.
    }
    \label{fig:nv-cost}
\end{figure}

The OU recursion requires $\mathcal O(MN)$ operations. The Kasdin convolution scales as $\mathcal O(MN\log N)$, while the
band-limited spectral synthesis scales as $\mathcal O(ML\log L)$ with $L\geq8N$.  Figure~\ref{fig:nv-cost} confirms the expected
dependence on record length and the approximately linear scaling with the number of independent trajectories. Absolute times are
hardware and implementation dependent and are shown only to compare the methods under common conditions.

\subsection{Reproducing the numerical results}
\label{sec:nv-repro}

The generators, estimators, and finite-record reference calculations are implemented in the open-source Julia package
\texttt{QuantumNoiseSimulator}. The OU update follows Ref.~\cite{nv-gillespie1996}, the band-limited synthesis follows
Refs.~\cite{nv-shinozuka1991,nv-timmer1995}, and the unrestricted power-law generator follows Ref.~\cite{kasdin1995}. The script
\texttt{paper/noise\_validation/make\_data.jl} writes the figure data and the file \texttt{numbers.tex}, from which the numerical
values in this section are imported directly. Fixed Xoshiro256++ seeds are stored with the scripts~\cite{nv-blackman2021}. The
qubit propagation was also cross-checked against QuTiP~\cite{nv-johansson2013}; with identical noise realizations and parameters,
the per-trajectory gate errors agreed to within $2\times10^{-8}$.

\end{document}